\documentclass[twocolumn]{aastex631}
\usepackage{showyourwork}
\usepackage{amsmath}
\usepackage{hyperref}

\shorttitle{Time-Variable Elemental Abundances in Loops}
\shortauthors{Reep et al.}

\begin{document}
 
\title{Modeling Time-Variable Elemental Abundances in Coronal Loop Simulations}

\author{Jeffrey W. Reep}
\affiliation{Institute for Astronomy, University of Hawai'i at M\=anoa, Pukalani, HI 96768}
\email{reep@hawaii.edu}

\author{John Unverferth}
\affiliation{National Research Council Postdoc at the Naval Research Laboratory, Washington, DC 20375}

\author{Will T. Barnes}
\affiliation{NASA Goddard Space Flight Center, Heliophysics Sciences Division, Greenbelt, MD 20771}
\affiliation{Department of Physics, American University, Washington, DC 20016}

\author{Sherry Chhabra}
\affiliation{George Mason University, Fairfax, VA, 22030}

\begin{abstract}
Numerous recent X-ray observations of coronal loops in both active regions (ARs) and solar flares have shown clearly that elemental abundances vary with time.  Over the course of a flare, they have been found to move from coronal values towards photospheric values near the flare peak, before slowly returning to coronal values during the gradual phase.  Coronal loop models typically assume that the elemental abundances are fixed, however.  In this work, we introduce a time-variable abundance factor into the 0D \texttt{ebtel++} code that models the changes due to chromospheric evaporation in order to understand how this affects coronal loop cooling.  We find that the for strong heating events ($\gtrsim$ 1 erg s$^{-1}$ cm$^{-3}$), the abundances quickly tend towards photospheric values.  For smaller heating rates, the abundances fall somewhere between coronal and photospheric values, causing the loop to cool more quickly than the time-fixed photospheric cases (typical flare simulations) and more slowly than time-fixed coronal cases (typical AR simulations).  This suggests heating rates in quiescent AR loops no larger than $\approx 0.1$ erg s$^{-1}$ cm$^{-3}$ to be consistent with recent measurements of abundance factors $f \gtrsim 2$.  
\end{abstract}


\section{Introduction}
\label{sec:intro}

Coronal loop models to date have fixed the elemental abundances both in space and time, using typically coronal abundances in quiescent loops or photospheric abundances in flare loops.  These abundances are used to calculate the bolometric radiative loss rate that directly impacts the cooling of the plasma.  Many recent X-ray observations of flares, however, clearly show that elemental abundances are not fixed and vary with time over the duration of the flare \citep{mondal2021, delzanna2021, mithun2022, nama2023, suarez2023, woods2023, rao2023, kepa2023}.  Similar observations have also found that the abundances of active regions (ARs) vary relatively quickly as well \citep{mondal2023}. 

There are two general mechanisms that could likely cause the change in elemental abundances: (1) preferential acceleration of ions, not neutrals, in the chromosphere, and (2) flows.  

The preferential acceleration of ions in the chromosphere is the cause of the first ionization potential effect (FIP effect), where Alfv\'en waves cause a ponderomotive force to accelerate ions from the chromosphere into the corona, while not affecting neutrals \citep{laming2015,laming2021}.  Hence, elements with low FIP ($\lesssim 10$ eV, \textit{e.g.} iron, calcium) that are ionized in the chromosphere are accelerated and become enhanced in coronal loops, while elements with high FIP ($\gtrsim 10$ eV, \textit{e.g.} helium, argon) are not, and their abundance generally does not vary with height.  Coronal loops within long-lived ARs typically show enhanced abundances of low-FIP elements (``coronal abundances'') relative to their values in the photosphere (``photospheric abundances'') as a result \citep{pottasch1964,mckenzie1992,delzanna2003,delzanna2014}.  

On the other hand, bulk flows between the chromosphere and corona carry both high- and low-FIP elements without preference.  Chromospheric evaporation, where a pressure expansion drives material from the chromosphere into the corona, must therefore cause a loop to fill with photospheric material.  We explore this effect in this work, for the first time modeling directly the change in elemental abundances due to chromospheric evaporation.  

\section{Modeling Time-Variable Abundance}

We use the \texttt{ebtel++} code\footnote{\url{https://github.com/rice-solar-physics/ebtelPlusPlus}, version 0.2 \url{https://zenodo.org/records/12675386}.} \citep{barnes2016} to model the hydrodynamics and the effects of time-variable elemental abundances in a coronal loop.  The code solves the zero-dimensional (coronally-averaged) hydrodynamics equations as a function of time for a coronal loop \citep{klimchuk2008,cargill2012a,cargill2012b}.  Because of the code's extreme speed, it can run parameter surveys within seconds, and generally compares well with higher dimensional models \citep{barnes2016}.  This makes it a powerful tool to understand the importance of several physical effects in various regimes.  

We have modified \texttt{ebtel++} to include time-variable abundances as follows.  We must first calculate how the abundance factor $f(t)$ changes as flows carry plasma into the corona.  Using that, we then incorporate it into the bolometric radiative loss rate, which dominates the cooling for much of a loop's lifetime.  The total radiative losses are given by $n^{2} \Lambda(T, n, f)$, where $\Lambda(T, n, f)$ is the radiative loss rate per unit emission measure.  To properly calculate the radiative losses, we must update $\Lambda(T, n, f)$ as each of the temperature $T$, density $n$, and abundance factor $f$ changes with time.

In the simulations here, we assume an initial abundance factor $f = 4$ in the corona, meaning that low-FIP elements like iron are enhanced by a factor of 4 over their photospheric values.  In comparison, observations of active regions and the solar wind find abundance factors ranging from around $f = 3$ to $f =5$ \citep{brooks2012}, with considerable spatiotemporal variation.  When the corona is heated, energy is transported to the chromosphere via thermal conduction, which causes the pressure in the chromosphere to rise and ablate material into the corona (chromospheric evaporation).  In the chromosphere, the plasma has photospheric composition with $f = 1$, so evaporation causes a shift in the proportion of coronal and photospheric material.  The abundance factor $f(t)$ in the corona can then be calculated as a weighted average between the initial material and that carried into the corona via evaporation:
\begin{align}
    f(t) &= \frac{f_{0} n_{0} + (n(t) - n_{0})}{n(t)}  \nonumber \\
     &= 1 + (f_{0} - 1) \frac{n_{0}}{n(t)}
\end{align}
\noindent where $n_{0} = n(t=0)$ and $f_{0} = f(t=0)$ are the initial coronal density and abundance factor, and $n(t)$ is the density in the corona as evaporation fills the loop.  There is one caveat to this equation: only upflows change the coronal abundance.  Unlike the ponderomotive force that likely causes the FIP effect, bulk flows do not cause preferential acceleration of different species.  However, since we have assumed that the corona initially has an abundance factor of $f = 4$ and the chromosphere $f = 1$, the evaporation that fills the loop naturally changes the composition.  When the loop is draining on the other hand, all elements drain in equal proportion, so the composition of the corona does not change.\footnote{A higher dimensional model might need to update the abundance factor at the top of the chromosphere, as the draining would slightly enhance the proportion of low FIP elements there.  However, this is likely to be an exceedingly small effect since the density of the chromosphere is significantly higher than the corona.}  We therefore only update $f(t)$ when the flows are \textit{into} the corona.  

We then modify the radiative loss function $\Lambda(T, n, f)$, which in previous works has generally only been treated as a function of temperature.  In the original implementation of \texttt{ebtel++} (and its IDL-based predecessor \texttt{EBTEL}), the loss function was parameterized as a power-law function in temperature \citep{klimchuk2008}.  Here, we have used the v10.1 of the CHIANTI atomic database \citep{dere1997,delzanna2021} to calculate look-up tables of $\Lambda(T, n, f)$ (using discrete steps in $T$, $n$, $f$).  Then, at each timestep, we use the current values of $T$, $n$, and $f$ to determine the loss rate with those tables.  We use the \citet{asplund2009} dataset for photospheric abundances ($f=1$).  As discussed in the appendix of \citet{reep2020}, for other values of $f > 1$, we enhance the abundances of the low-FIP elements (\textit{e.g.} for $f=4$, the abundance of iron is 4 times greater than in the \citealt{asplund2009} dataset).  

\section{Loop Simulations}
We now examine the effect of time-variable abundance on the hydrodynamics of coronal loops.  We examine loops of lengths 40 and 80 Mm, with various heating rates.  We compare the hydrodynamic evolution of these loops when assuming time-fixed photospheric abundances, time-fixed coronal abundances, and time-variable abundances. 

\begin{figure*}
    \script{render_figure1.py}
    \begin{centering}
    \includegraphics[width=0.32\linewidth]{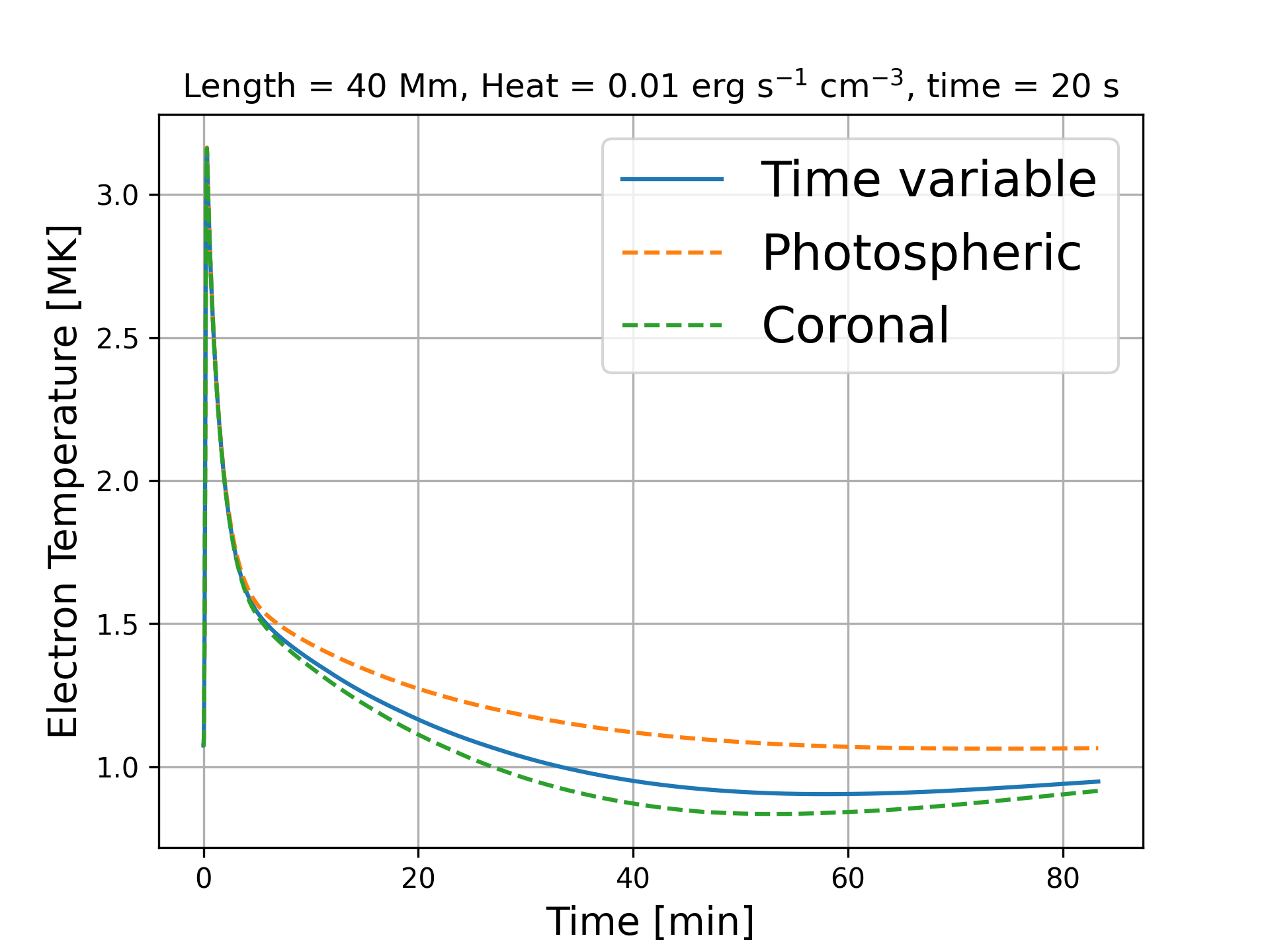}
    \includegraphics[width=0.32\linewidth]{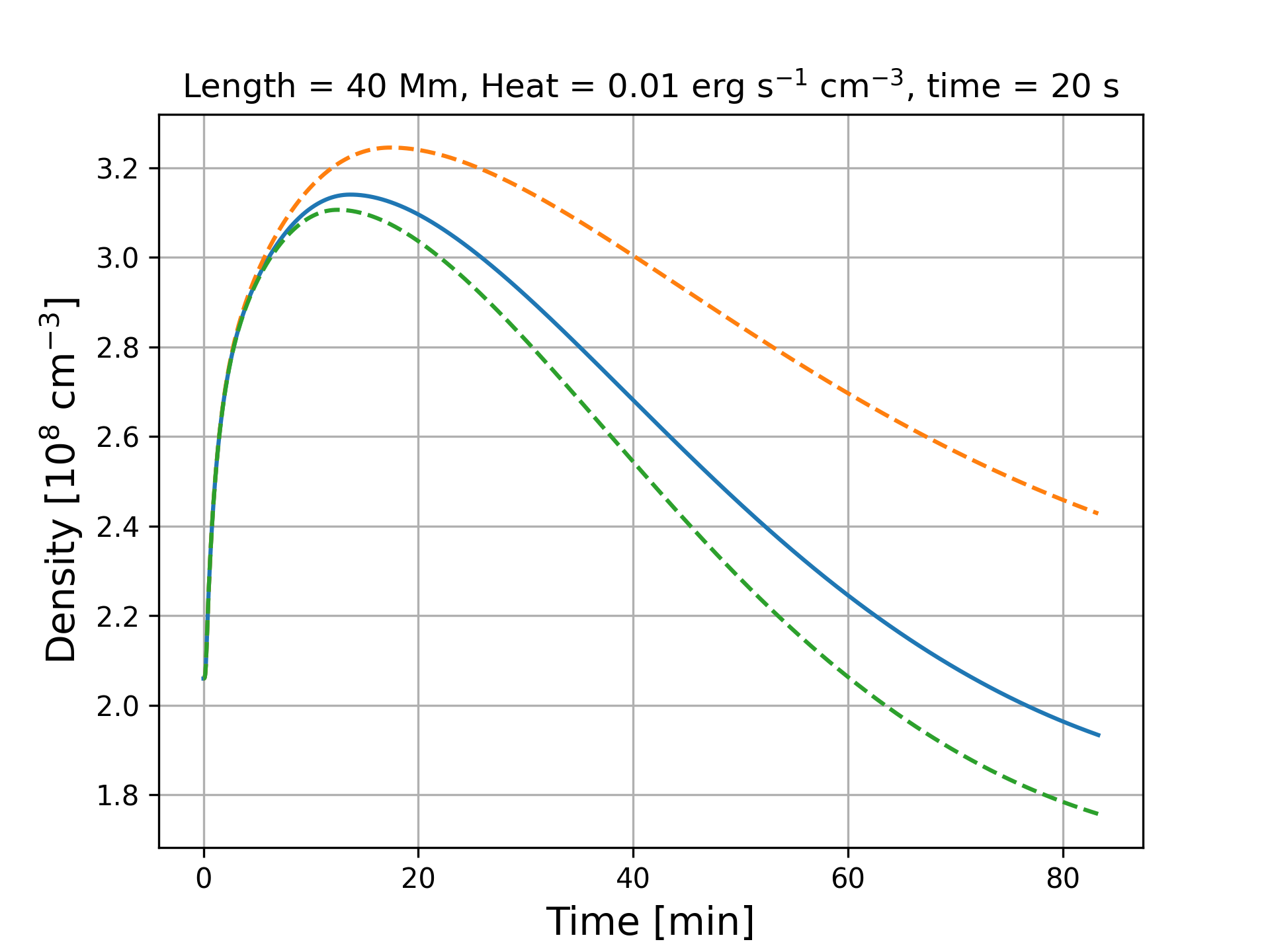}
    \includegraphics[width=0.32\linewidth]{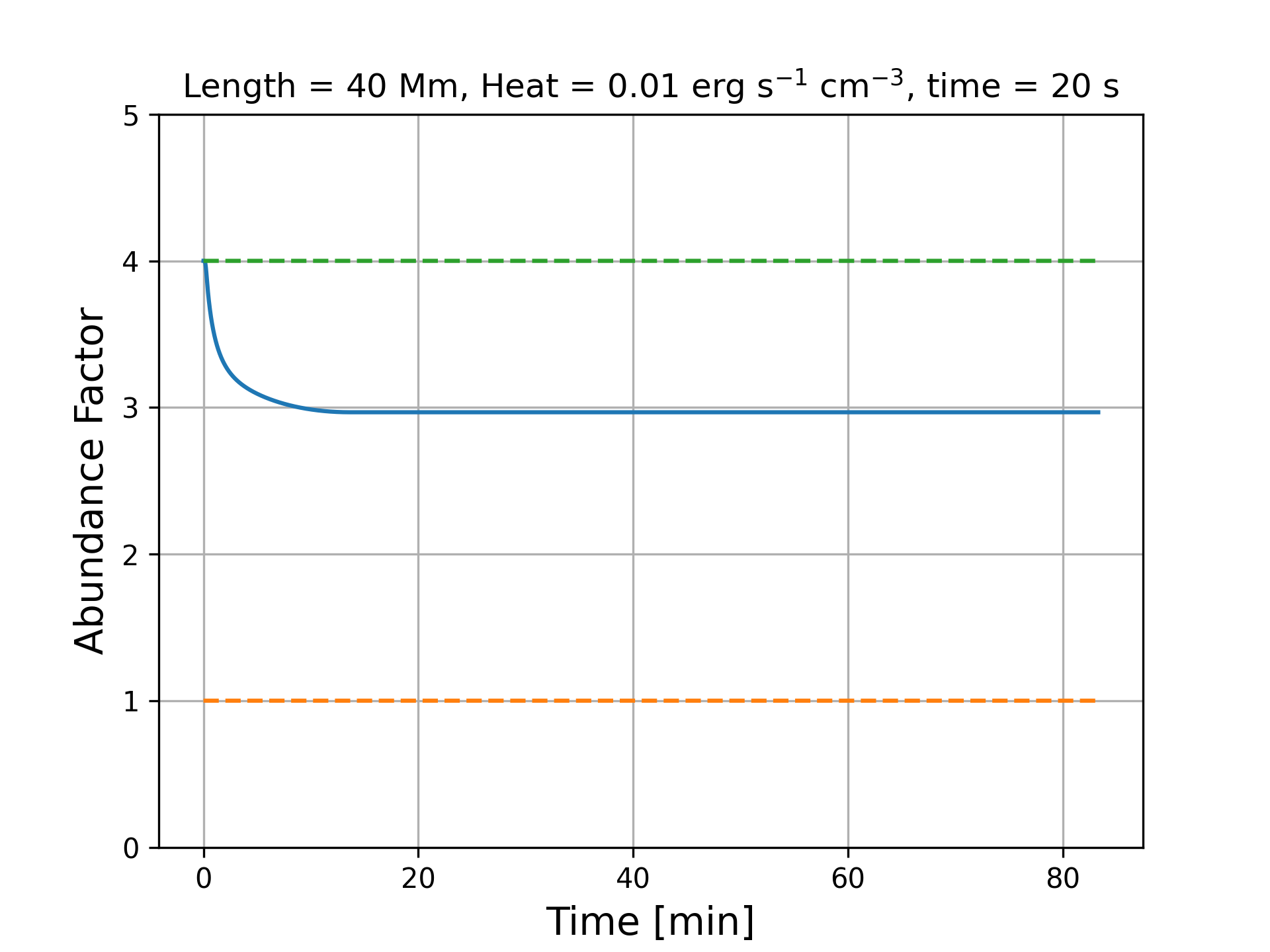}
    \includegraphics[width=0.32\linewidth]{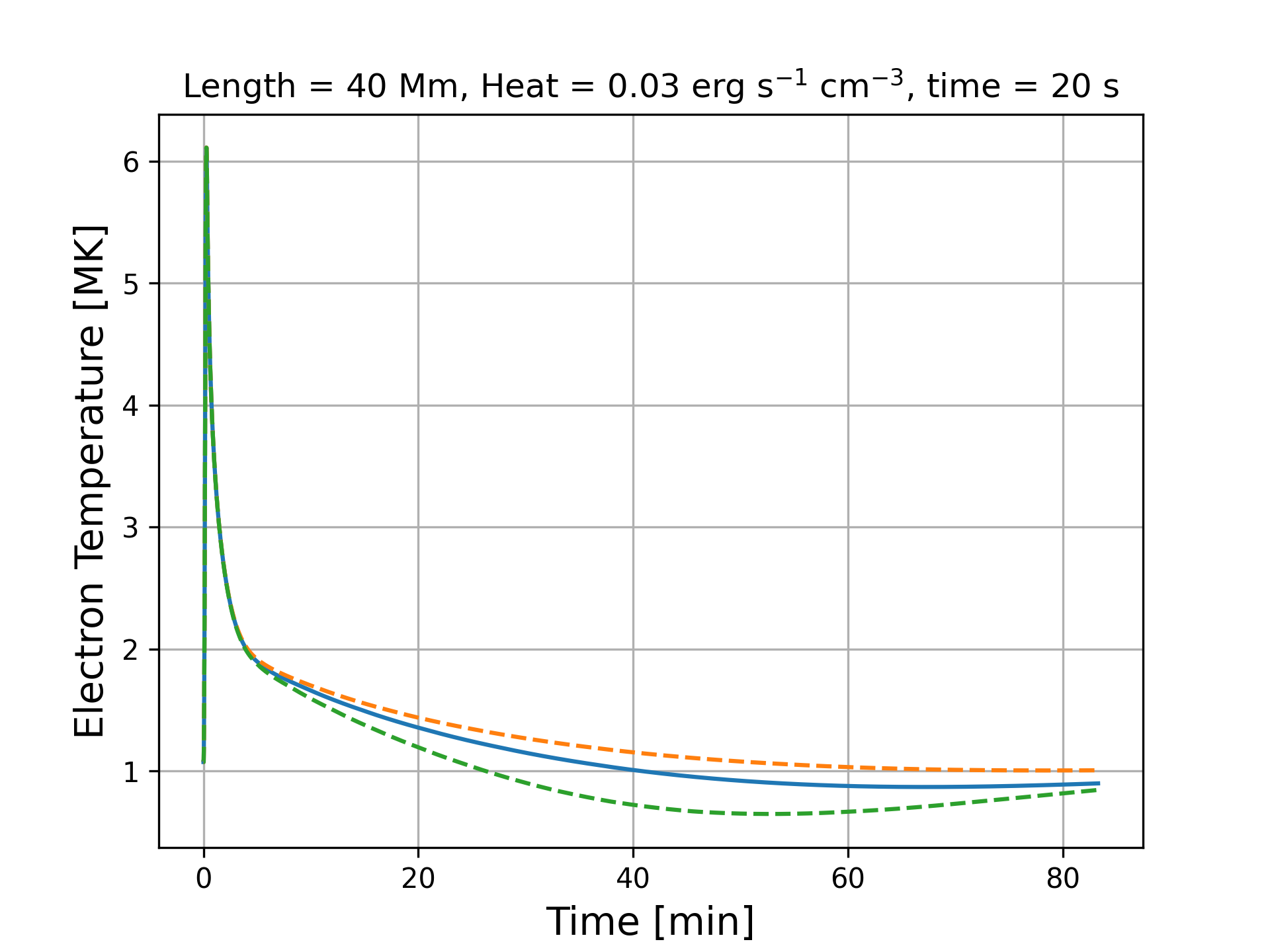}
    \includegraphics[width=0.32\linewidth]{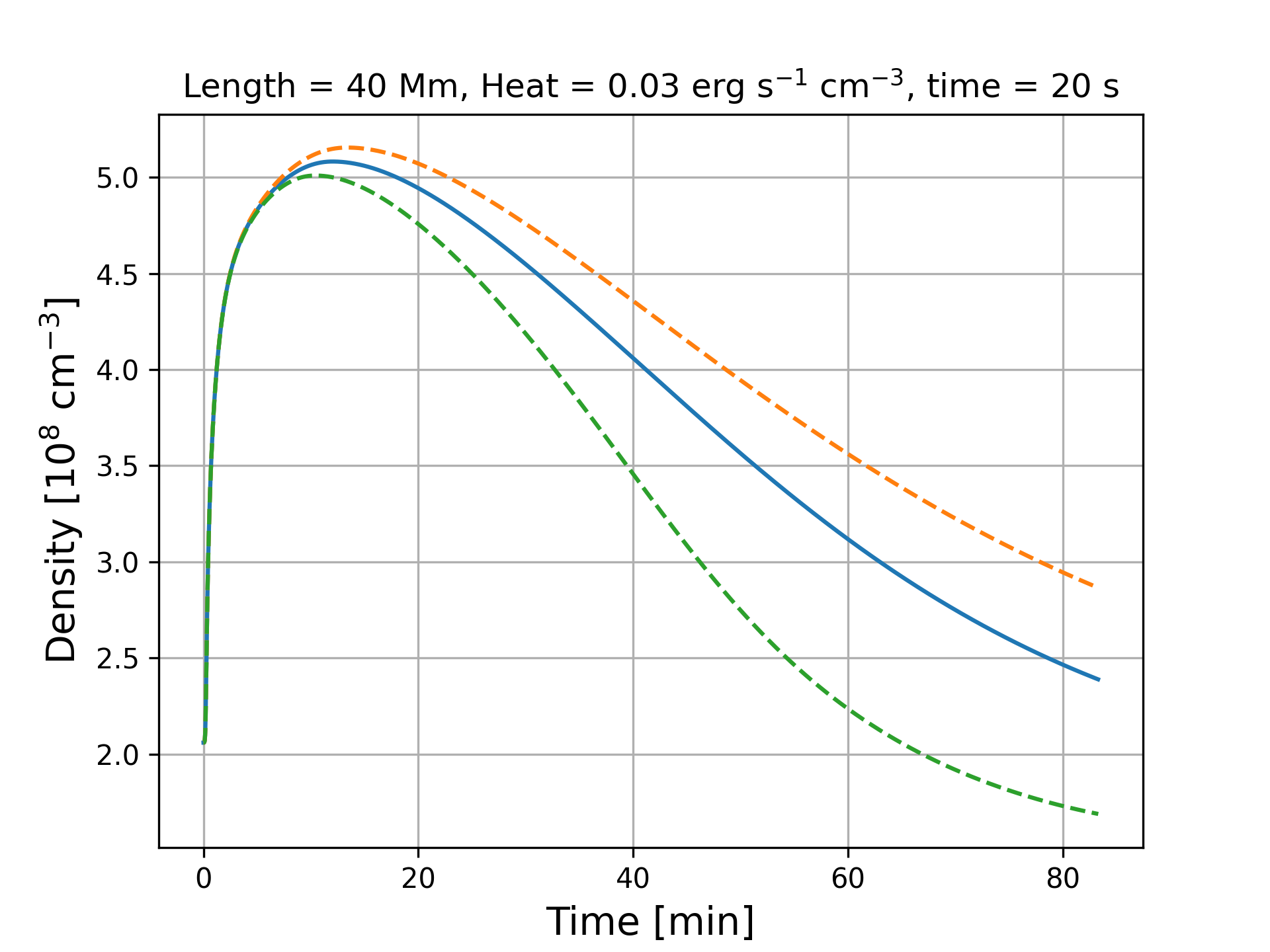}
    \includegraphics[width=0.32\linewidth]{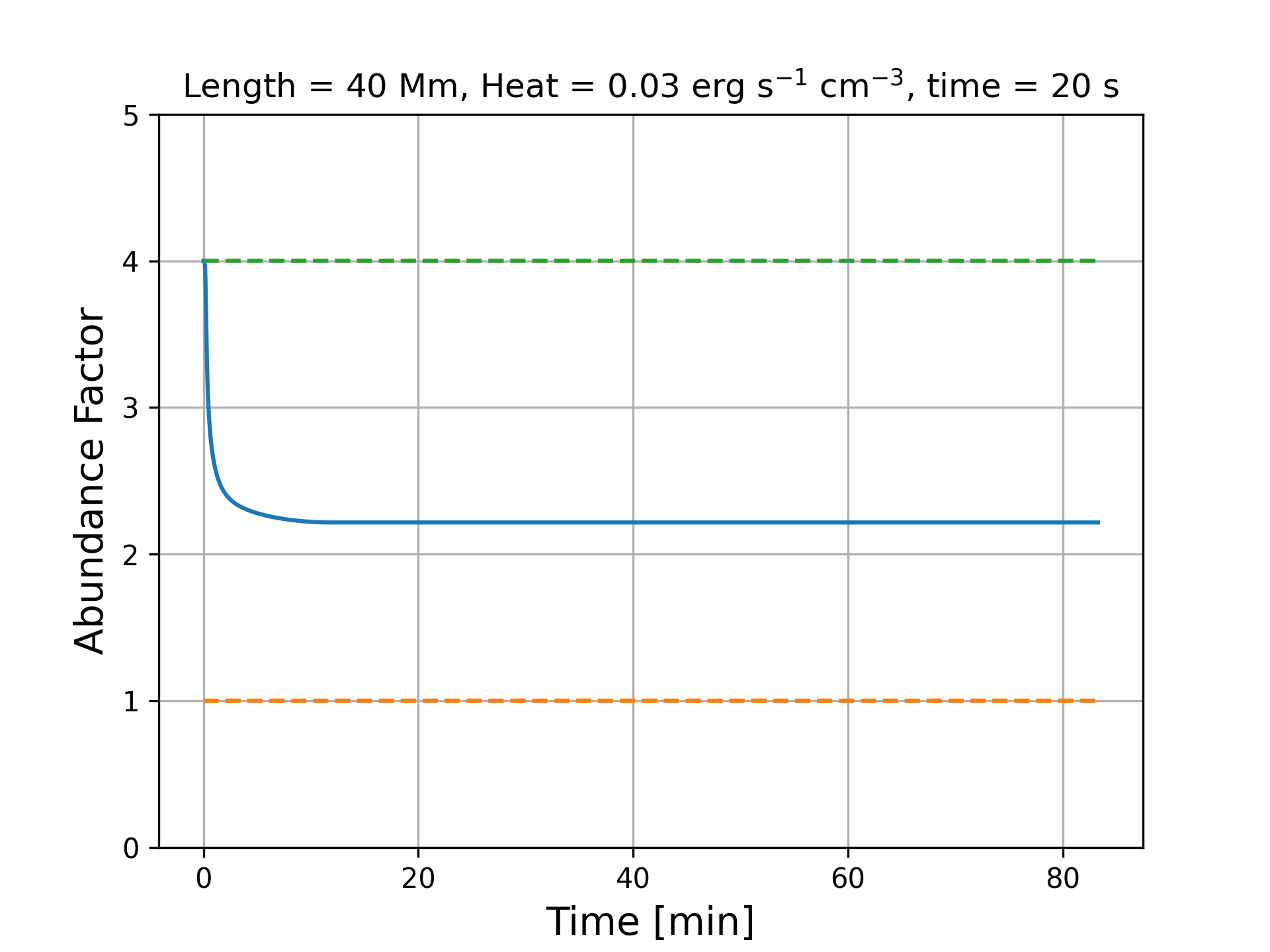}    \includegraphics[width=0.32\linewidth]{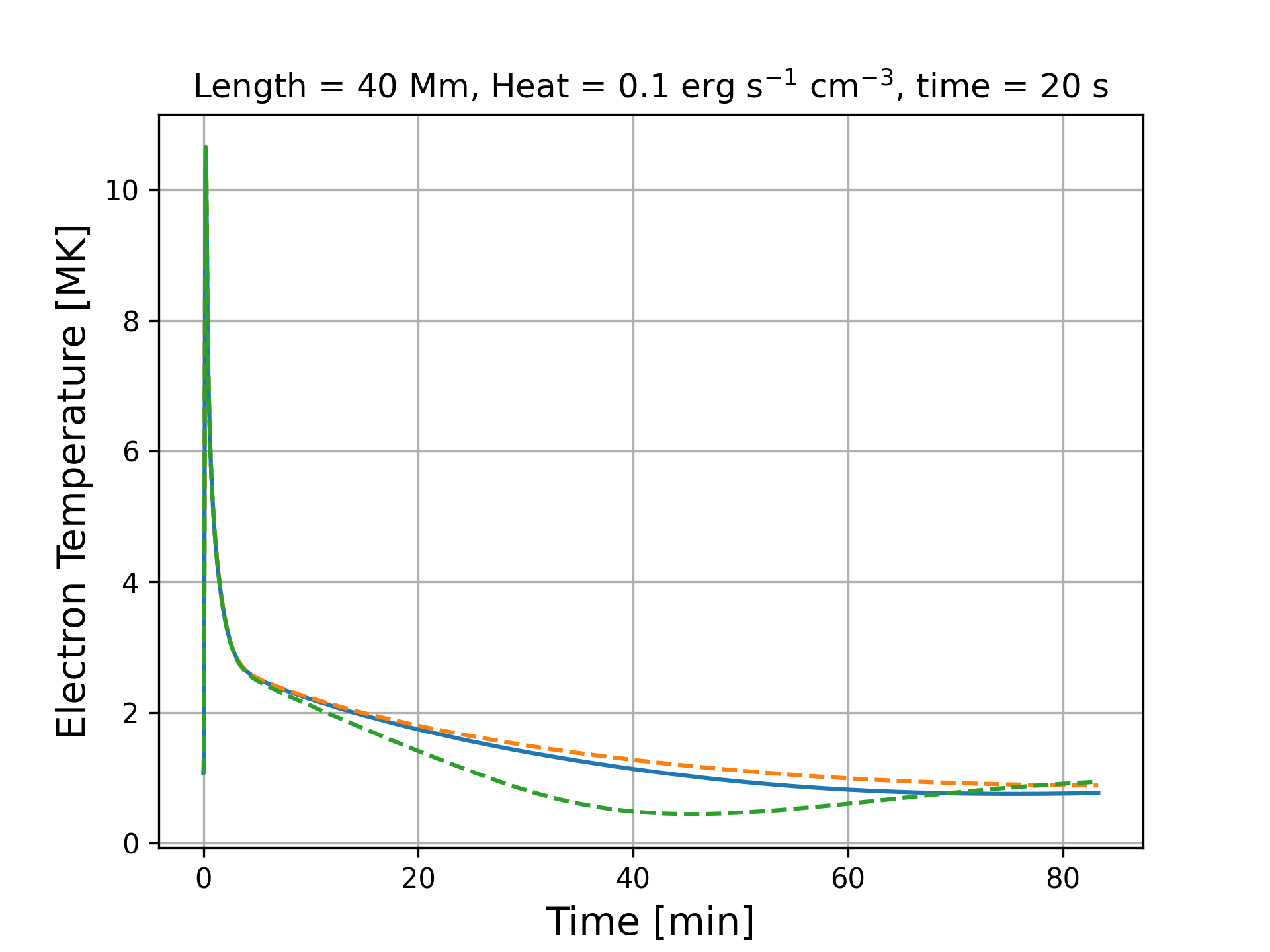}
    \includegraphics[width=0.32\linewidth]{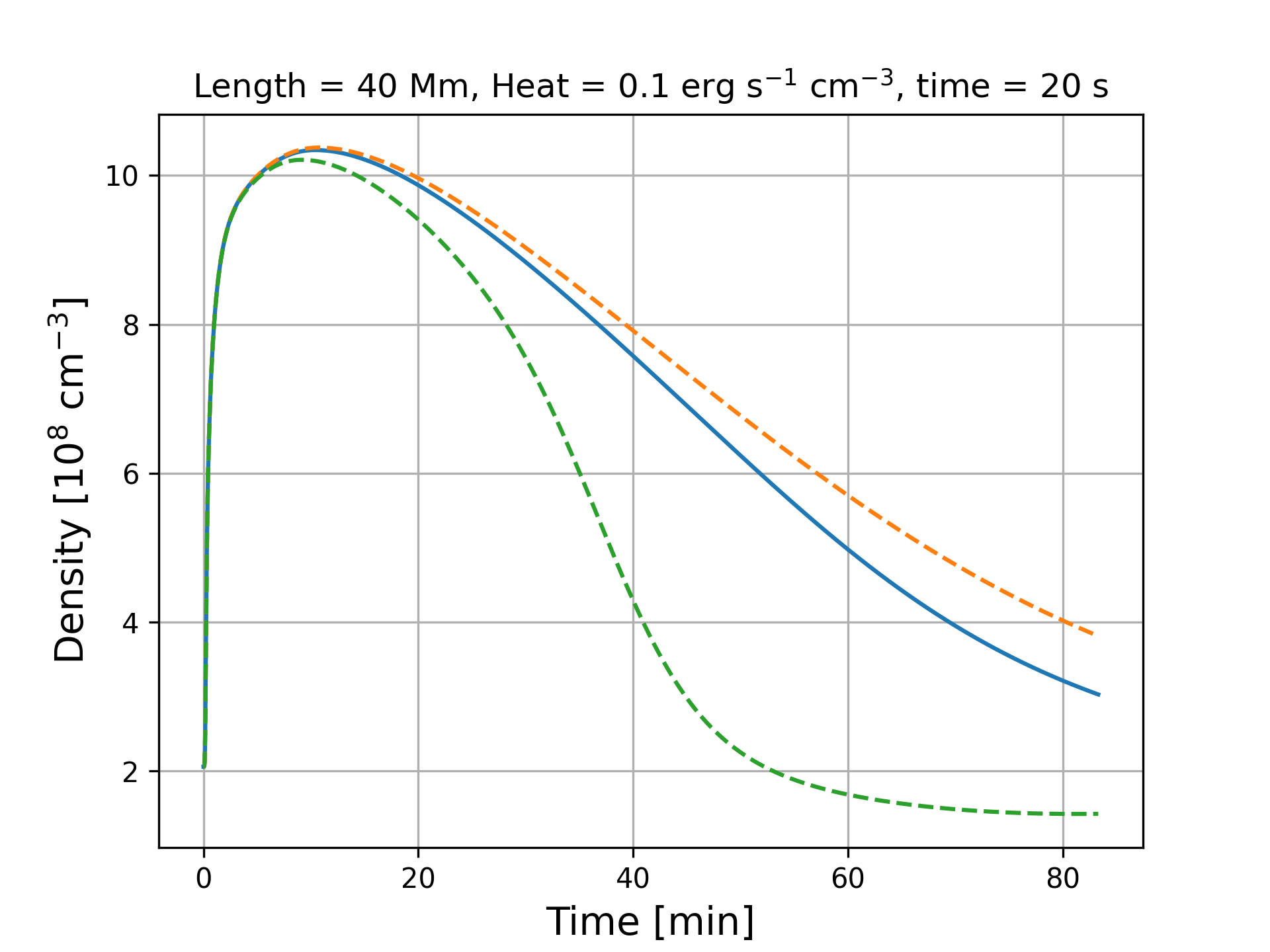}
    \includegraphics[width=0.32\linewidth]{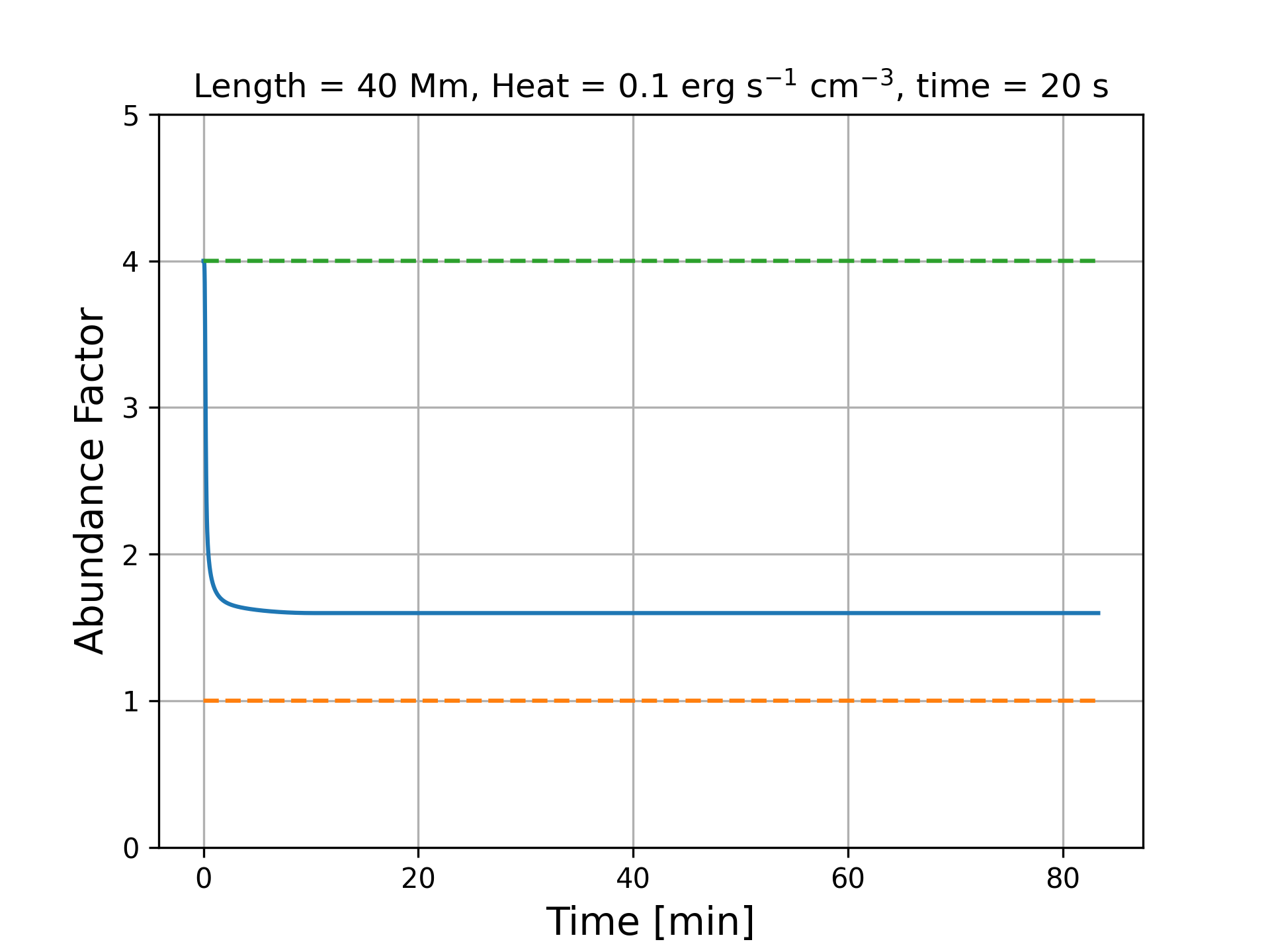}
    \includegraphics[width=0.32\linewidth]{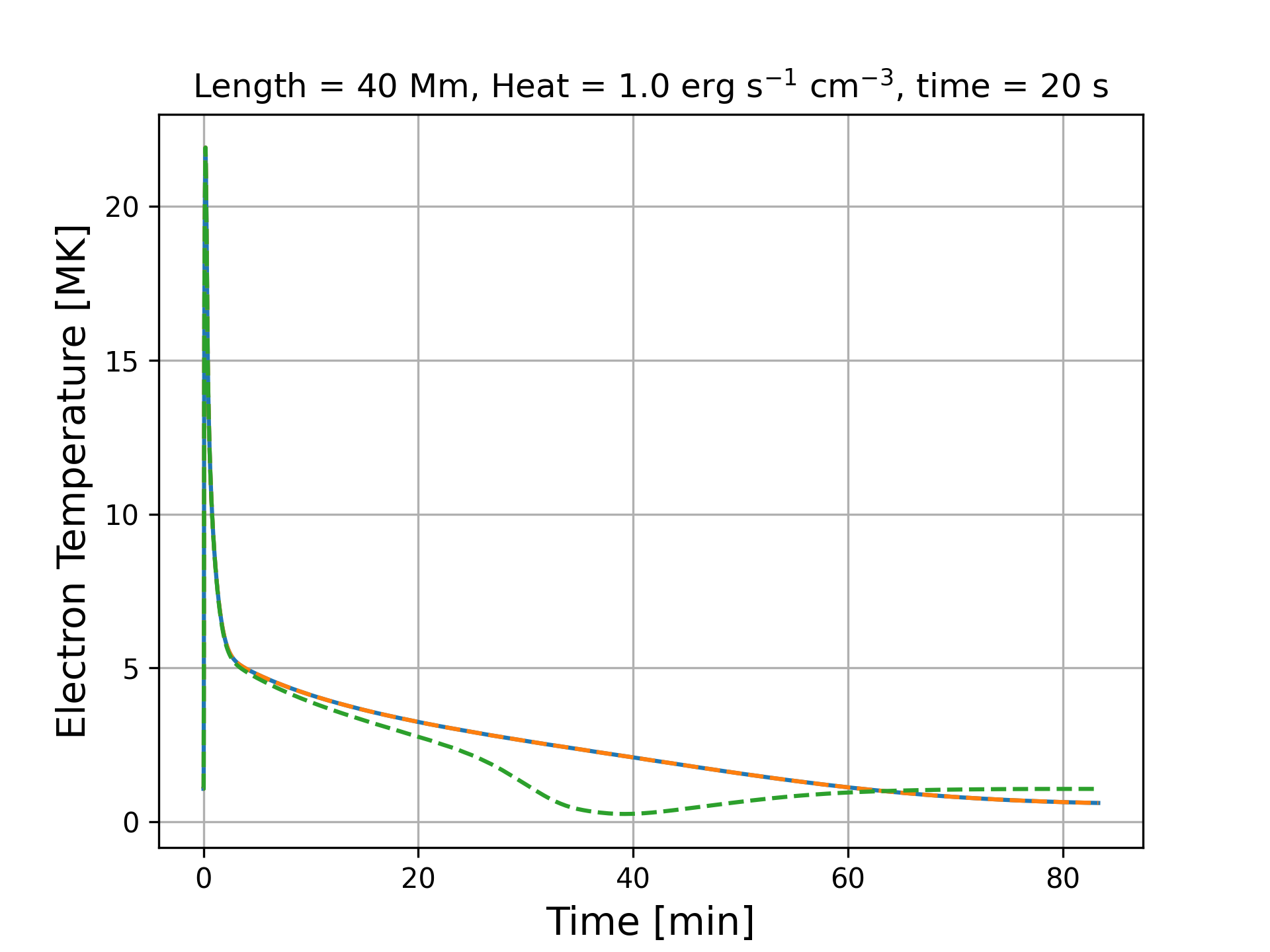}
    \includegraphics[width=0.32\linewidth]{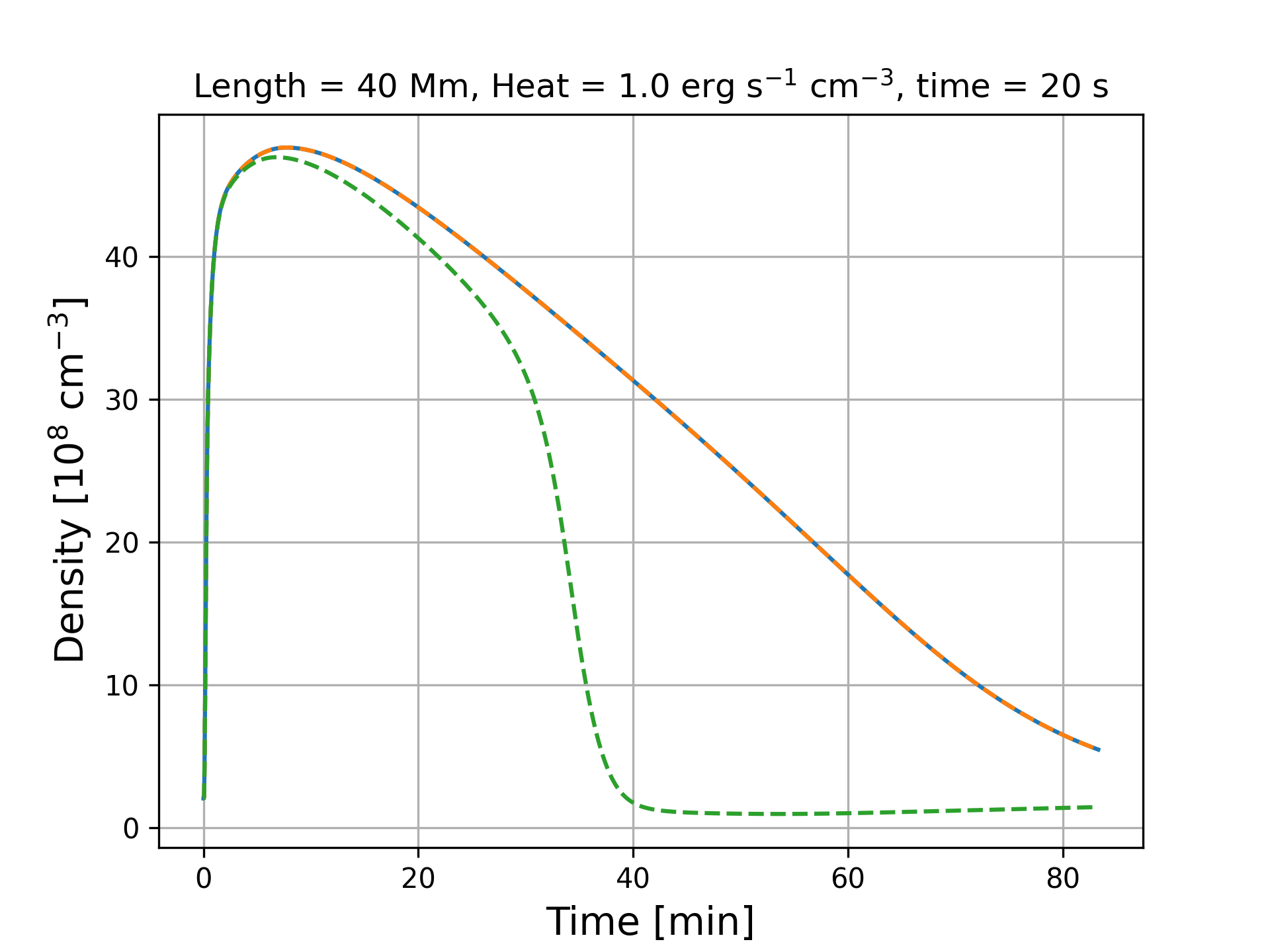}
    \includegraphics[width=0.32\linewidth]{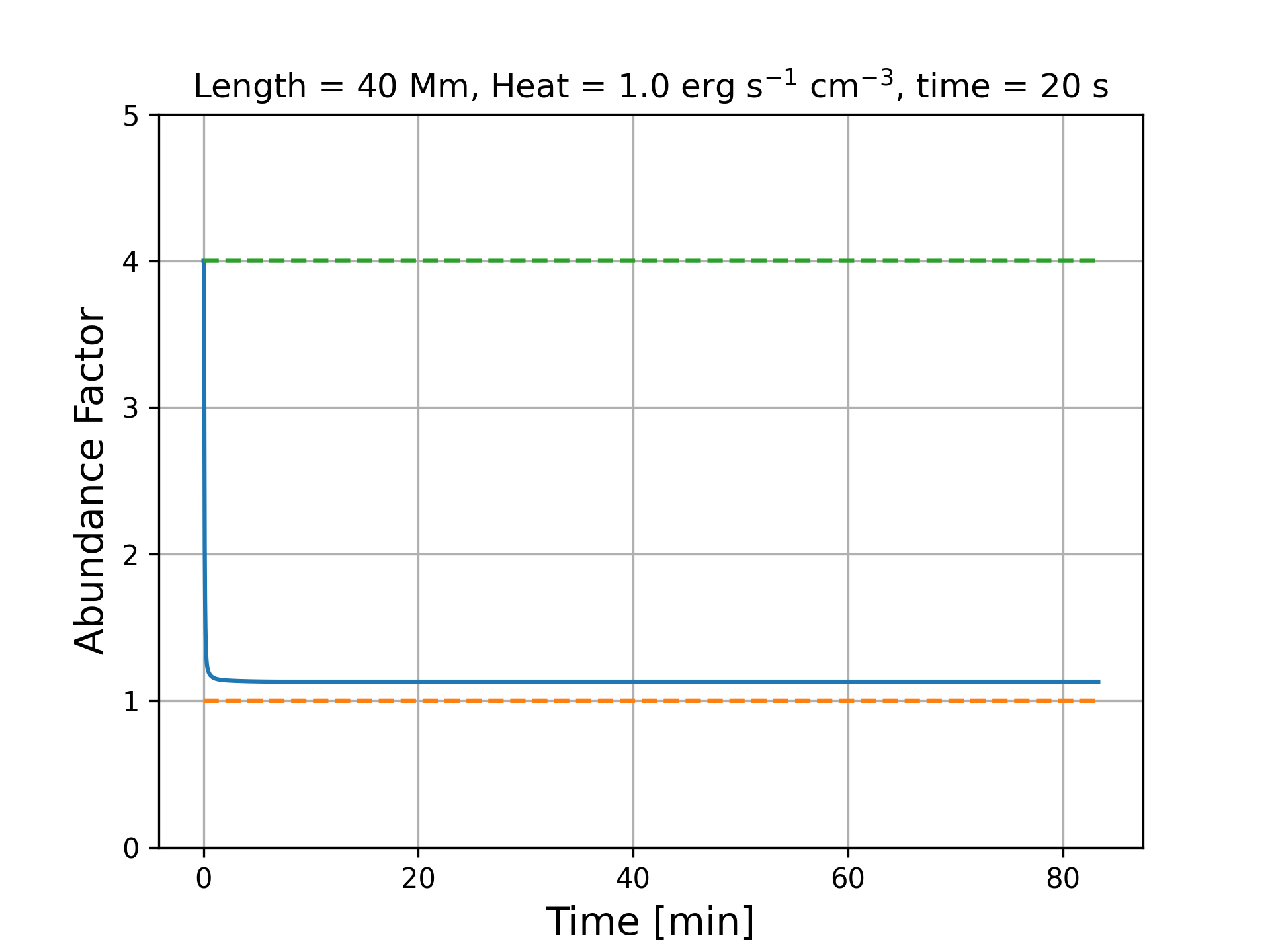}
    \caption{The evolution of the electron temperature (left column), density (center column), and abundance factor (right column) for a 40 Mm coronal loop, heated with 0.01, 0.03, 0.1, and 1.0 erg s$^{-1}$ cm$^{-3}$ (rows), for a 20 s heating pulse.  The blue lines show the case with time-variable abundances $f(t)$, while orange lines show time-fixed photospheric abundances ($f=1$), and green time-fixed coronal abundances ($f=4$). }
    \label{fig:L40}
    \end{centering}
\end{figure*}

Figure \ref{fig:L40} shows a comparison of evolution for 40 Mm loops, with the electrons heated by a 20 s heating pulse on a triangular profile (10 s rise, 10 s decay), for maximum heating rates of 0.01, 0.03, 0.1, and 1.0 erg s$^{-1}$ cm$^{-3}$ (each row, respectively).  The columns respectively show the evolution of the electron temperature $T_{e}(t)$, density $n(t)$, and abundance factor $f(t)$.  The different lines show the three cases of abundances: time-variable abundance factor (blue), time-fixed photospheric abundances ($f=1$, orange), and time-fixed coronal abundances ($f=4$, green).  

It is clear that the heating rates, and thus the strength of chromospheric evaporation, play a large role in the evolution of the abundance factor and therefore the radiative losses.  Since radiative losses are stronger with coronal abundances than with photospheric ones, this causes the time-variable case to cool faster than the photospheric case but slower than the coronal case in general.  In the weakest heating case, the coronal density increases by around 50\%, causing the abundance factor to fall to 3 over the course of 10 minutes, compared to the strongest heating case where the abundance factor falls to photospheric levels almost immediately.  Notably, since radiation is relatively weak prior to the end of chromospheric evaporation, the temperatures and densities of the impulsive phase are relatively unaffected by the abundance values.  The cases only start to diverge in the cooling period, after radiation becomes the dominant cooling mechanism.  We note that short heating pulses in \texttt{ebtel++} (like those used here) overestimate the density relative to higher dimensional models by around 20\% because of the lack of spatial extent \citep{barnes2016}, which may exacerbate the divergences.

\begin{figure*}
    \script{render_figure2.py}
    \centering
    \includegraphics[width=0.32\linewidth]{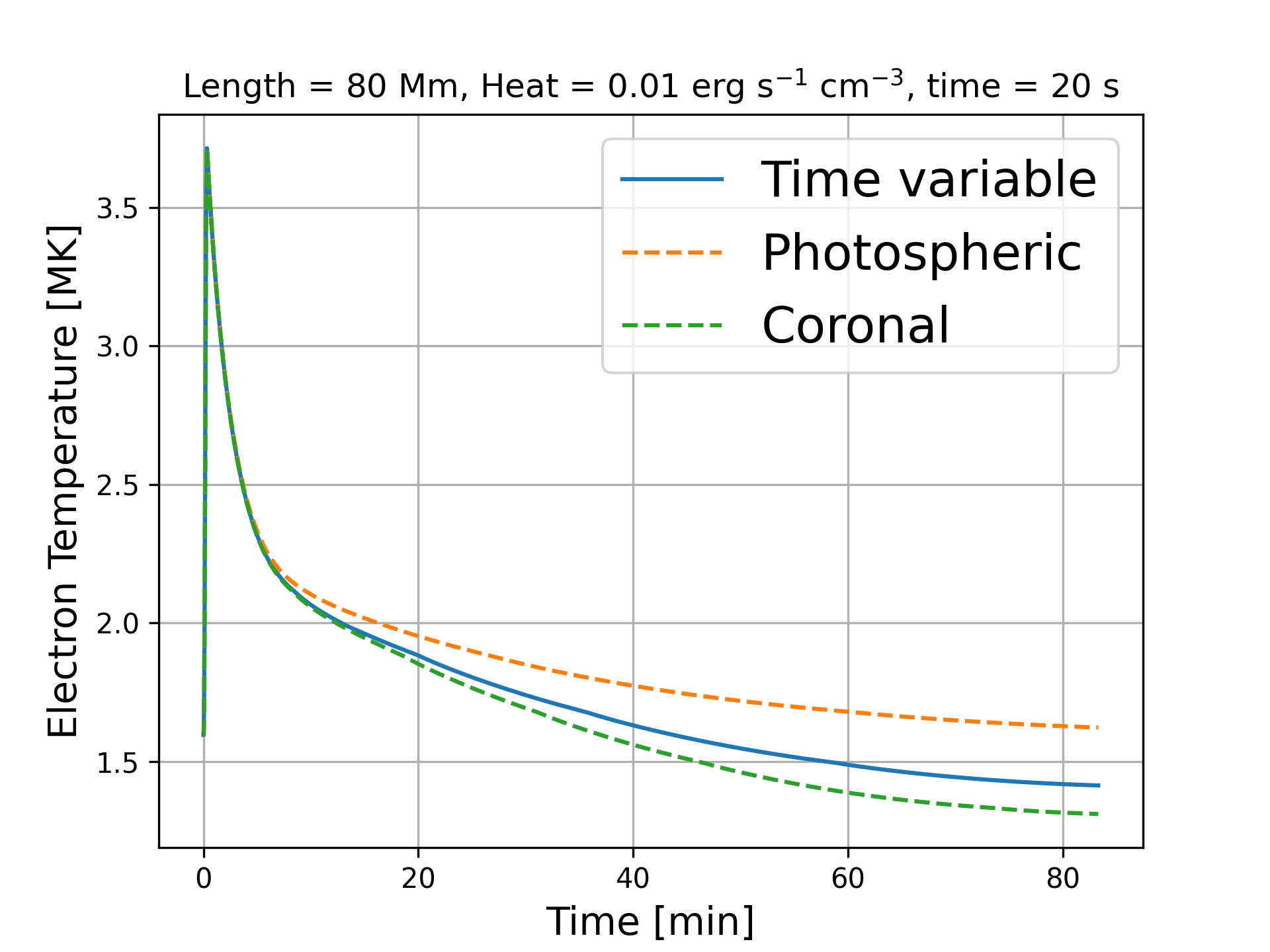}
    \includegraphics[width=0.32\linewidth]{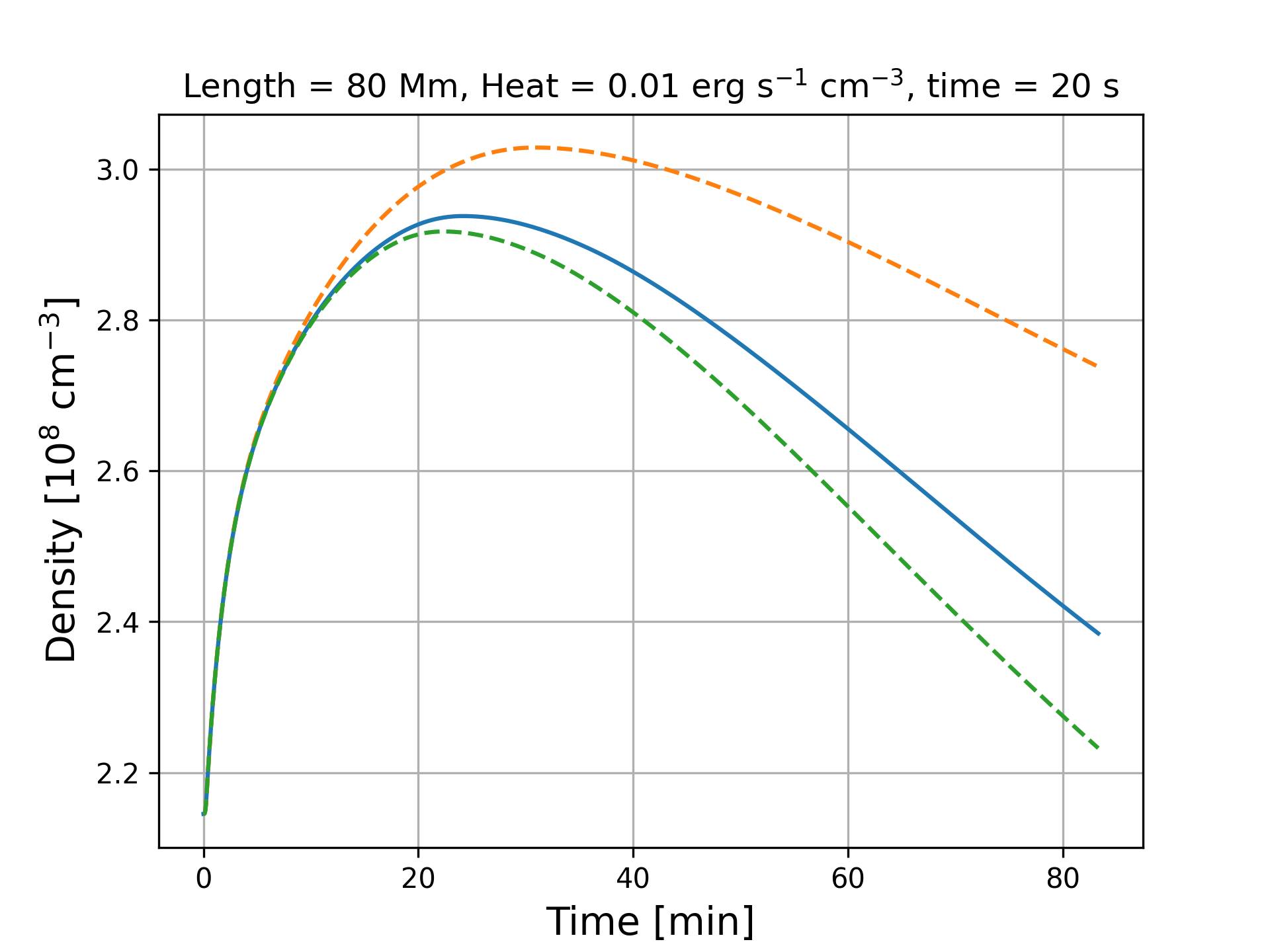}
    \includegraphics[width=0.32\linewidth]{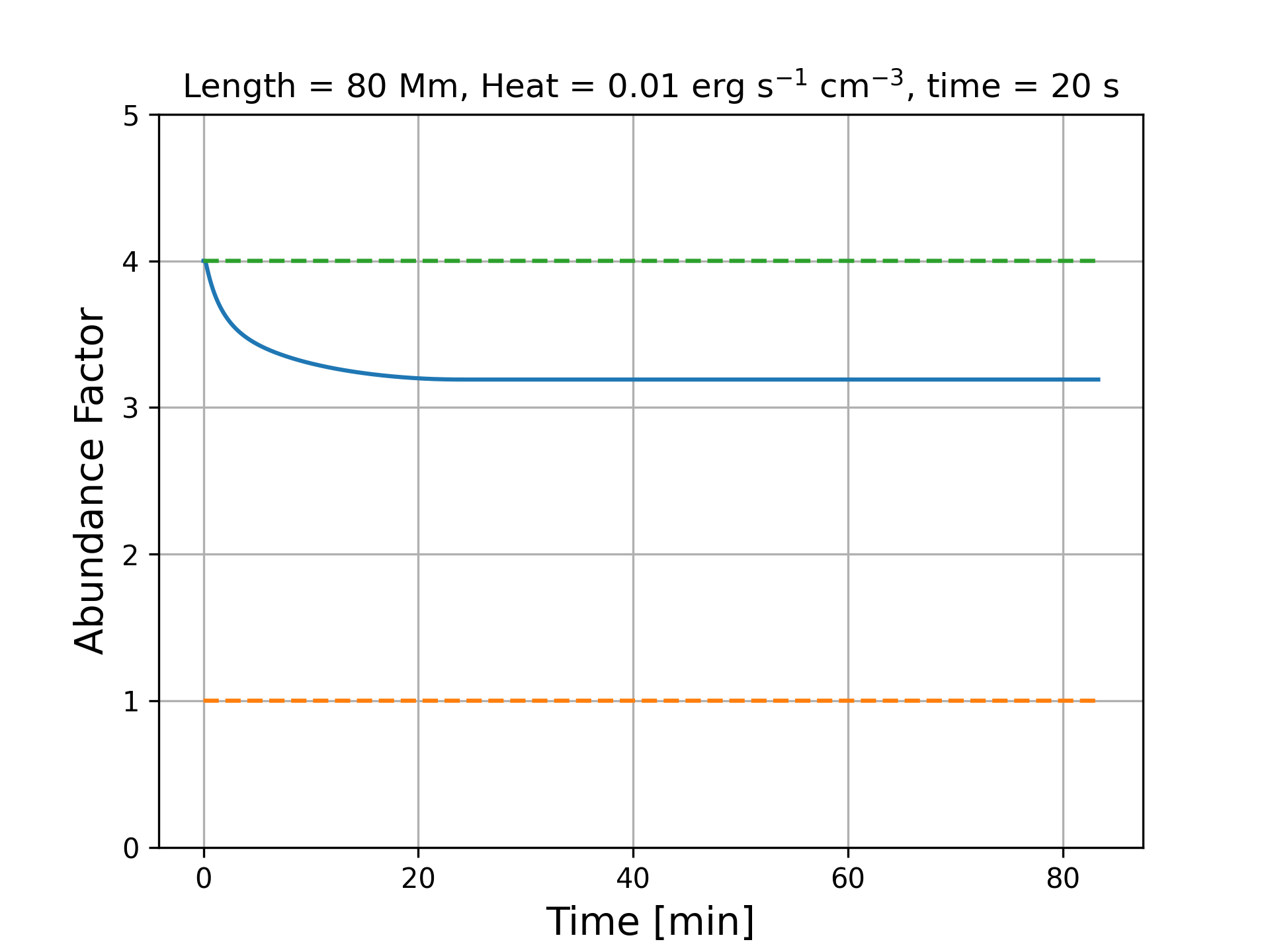}
    \includegraphics[width=0.32\linewidth]{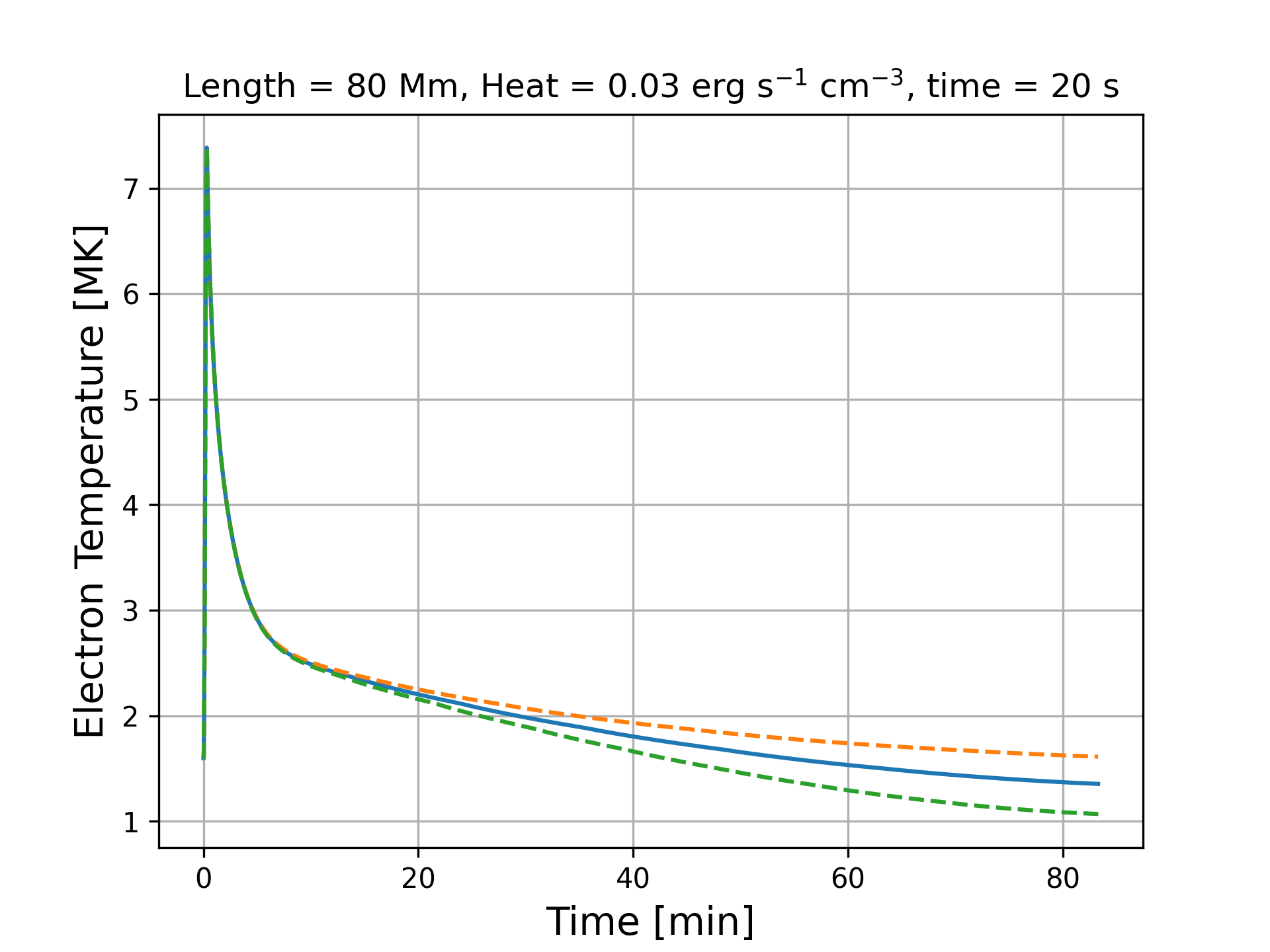}
    \includegraphics[width=0.32\linewidth]{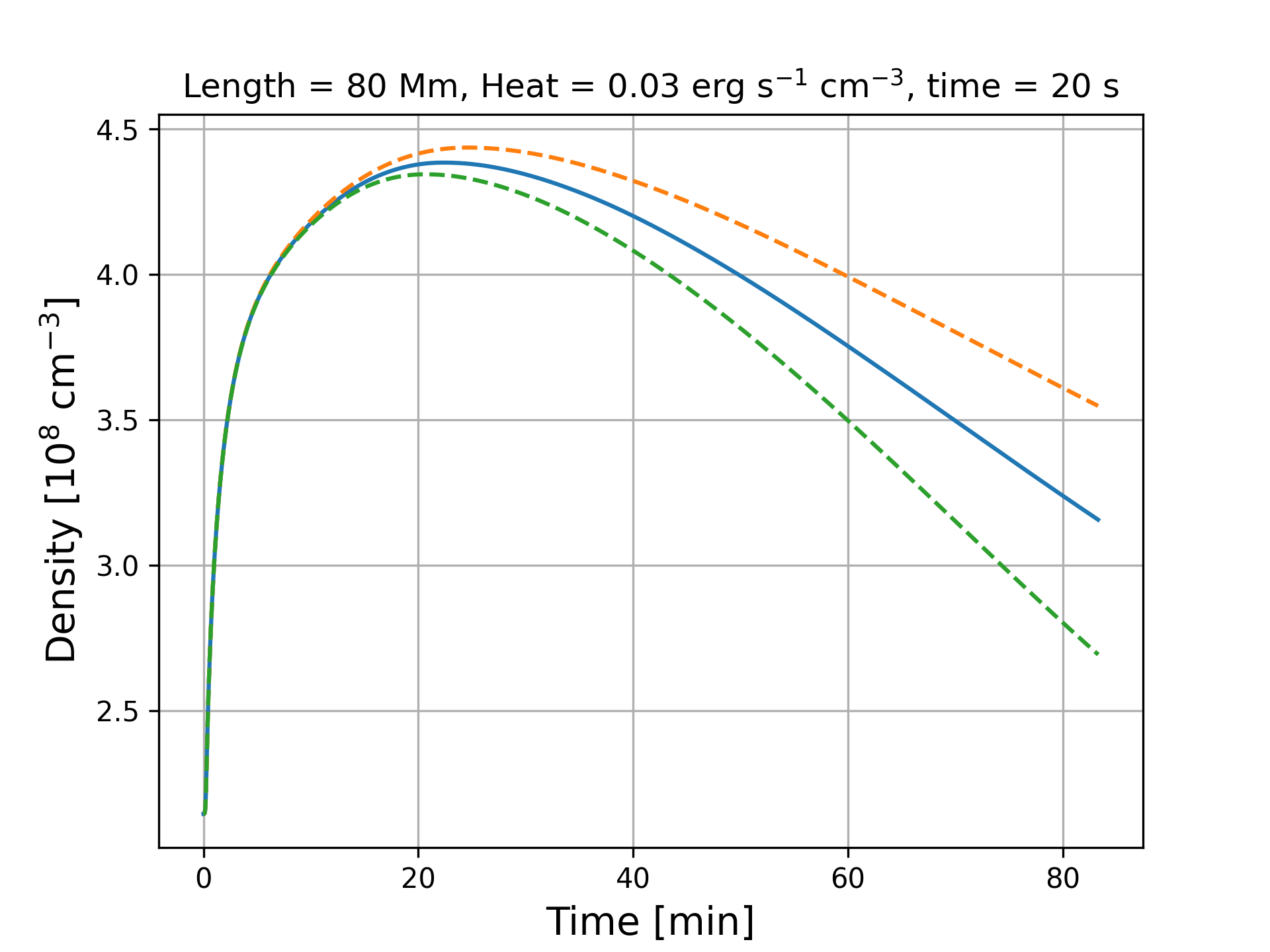}
    \includegraphics[width=0.32\linewidth]{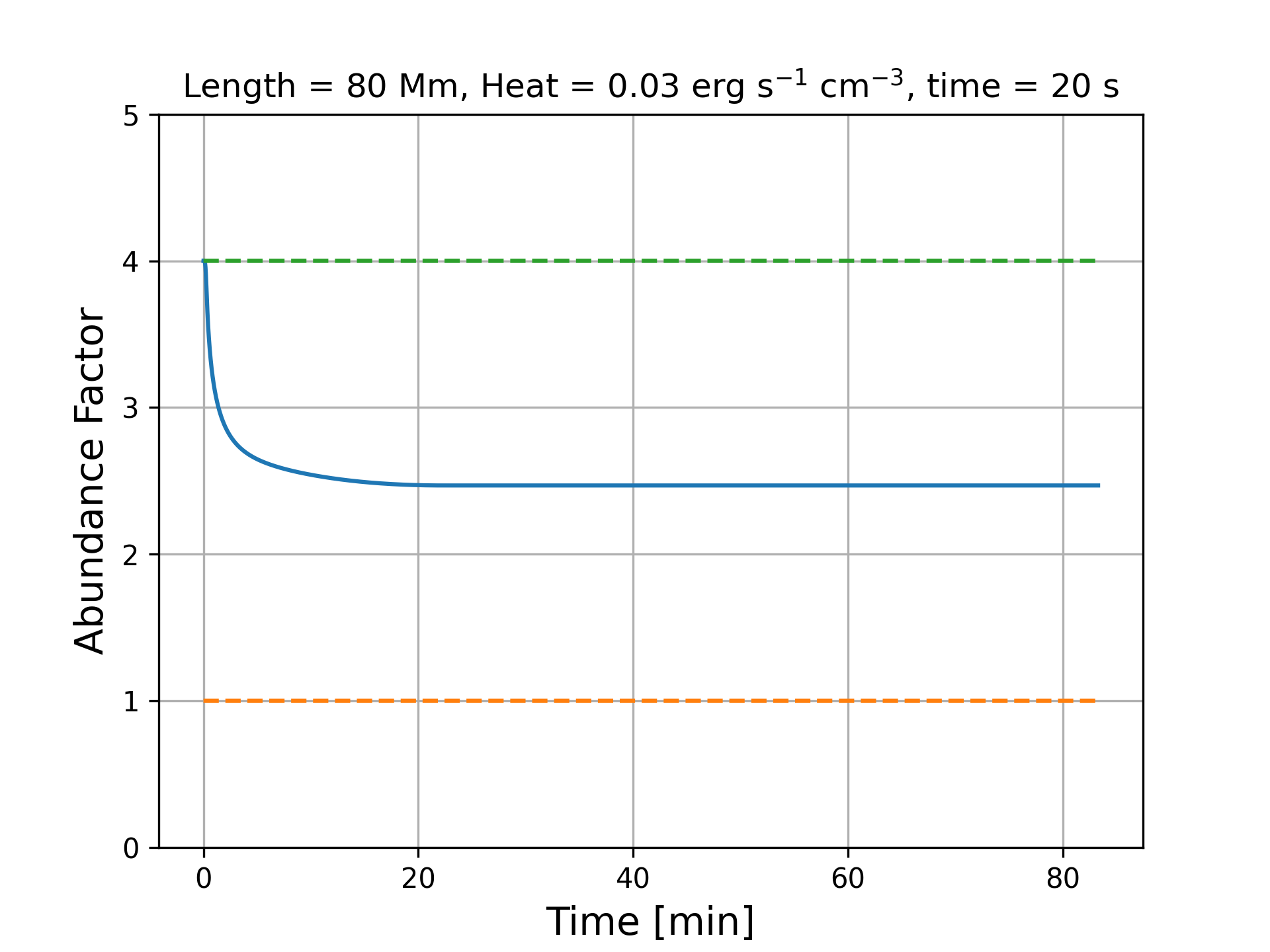}
    \includegraphics[width=0.32\linewidth]{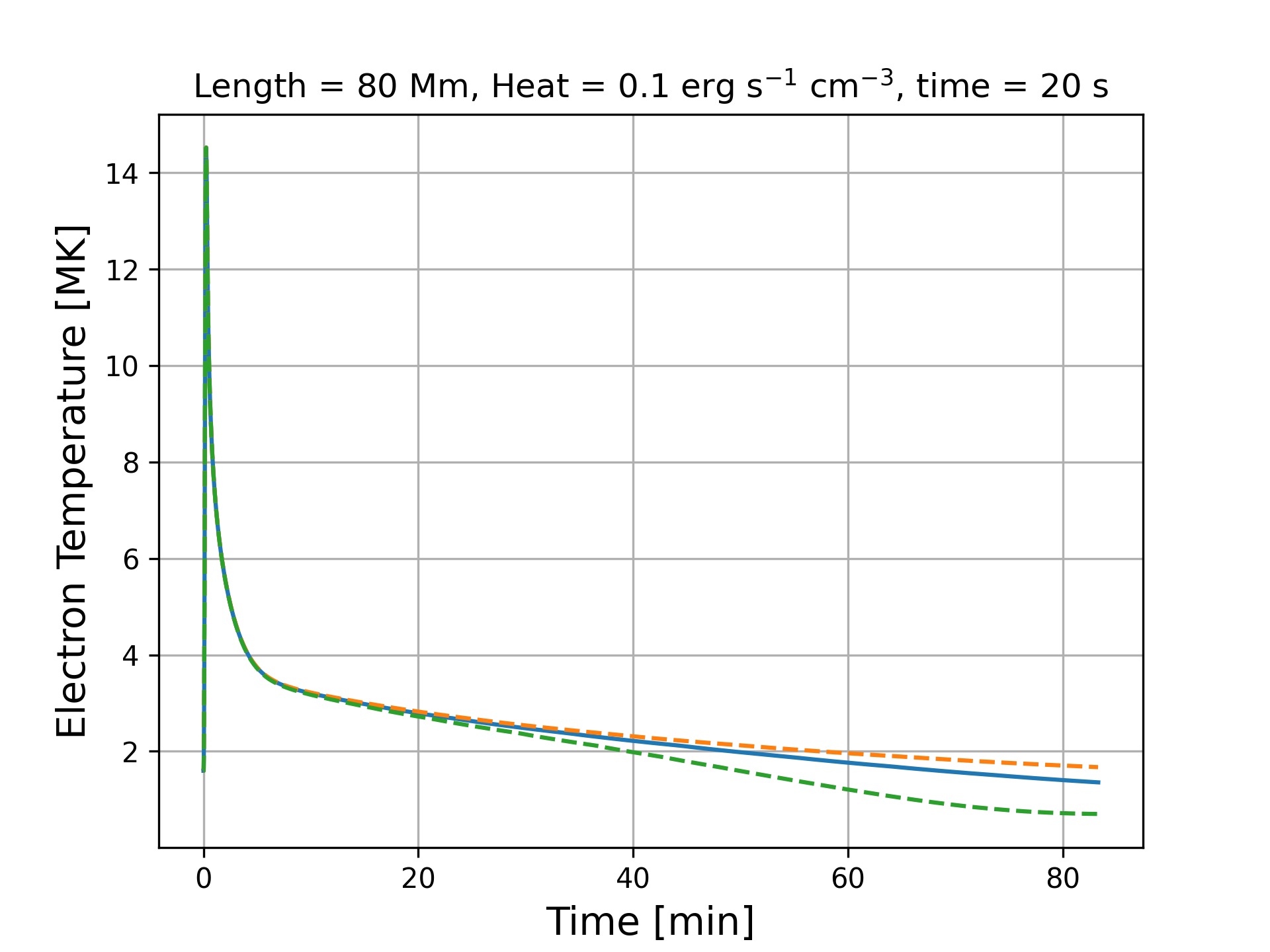}
    \includegraphics[width=0.32\linewidth]{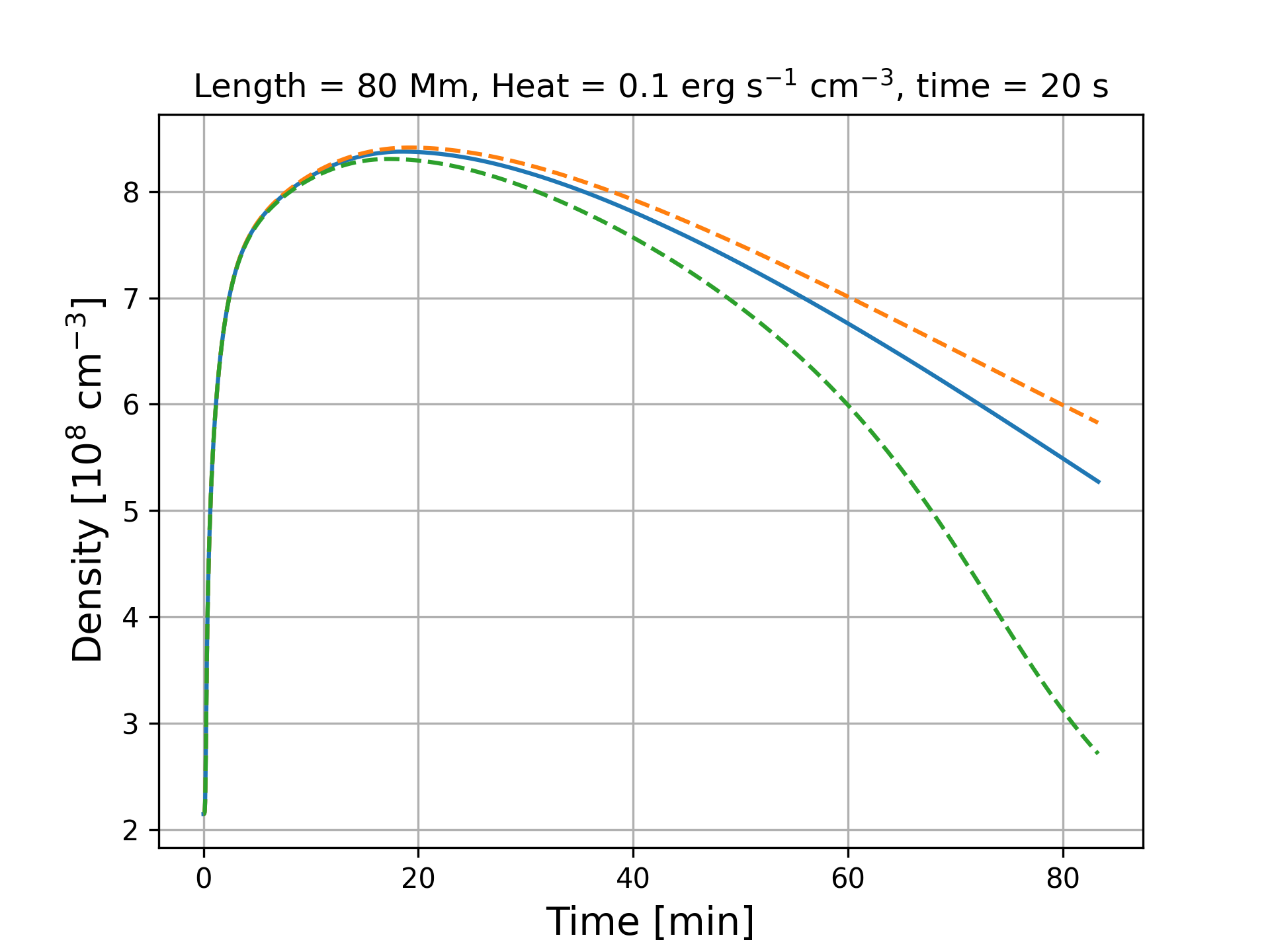}
    \includegraphics[width=0.32\linewidth]{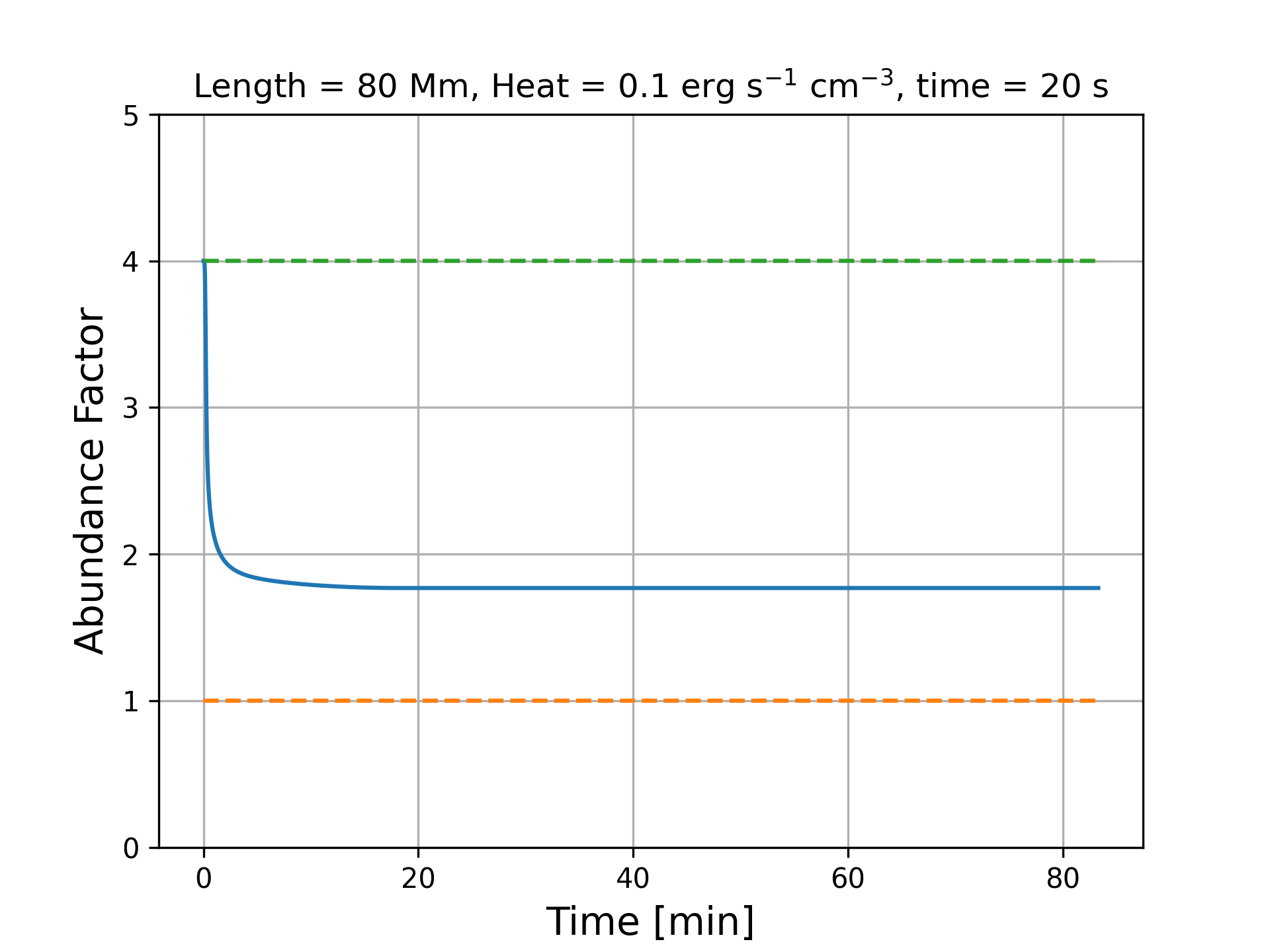}
    \includegraphics[width=0.32\linewidth]{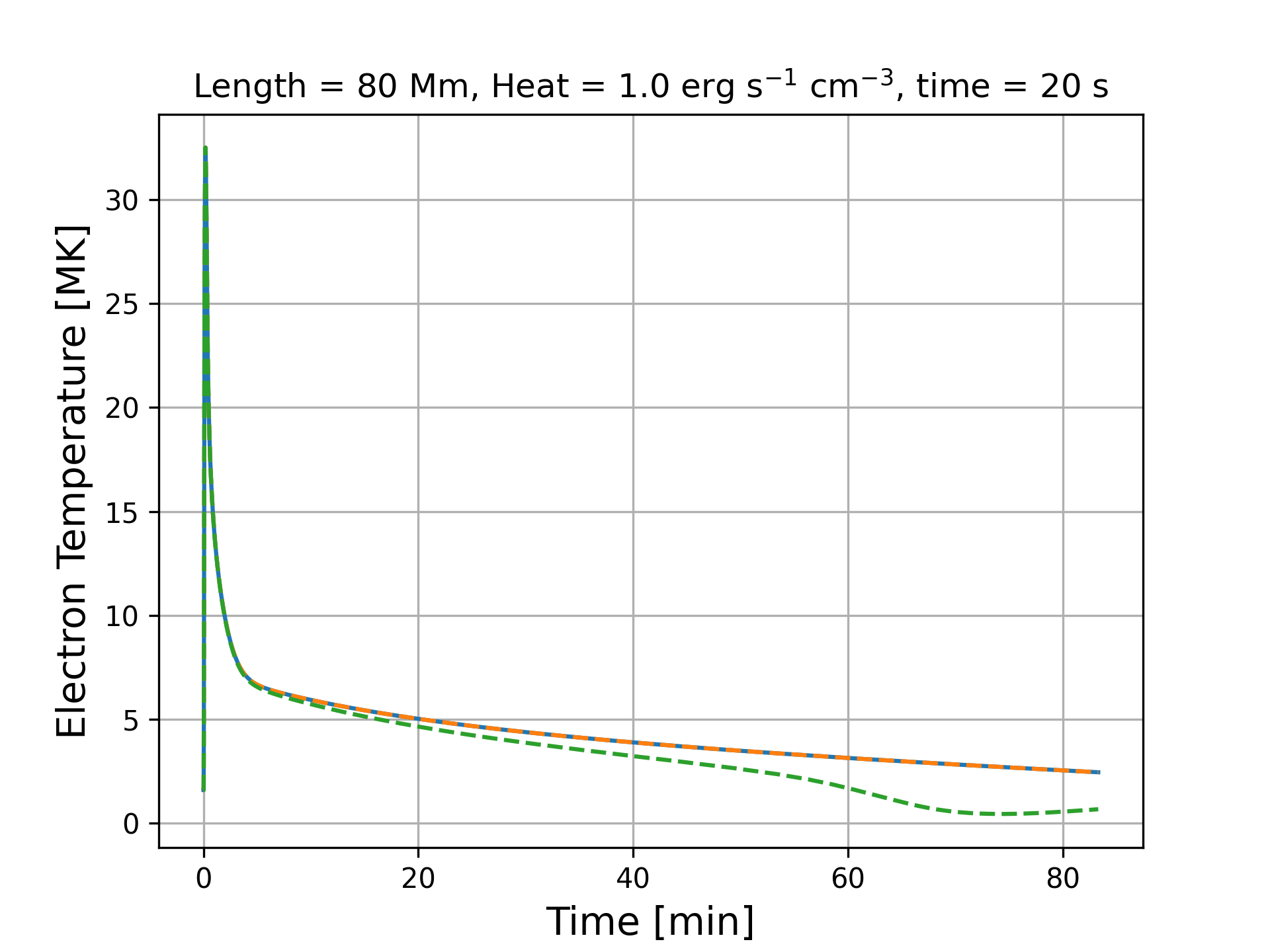}
    \includegraphics[width=0.32\linewidth]{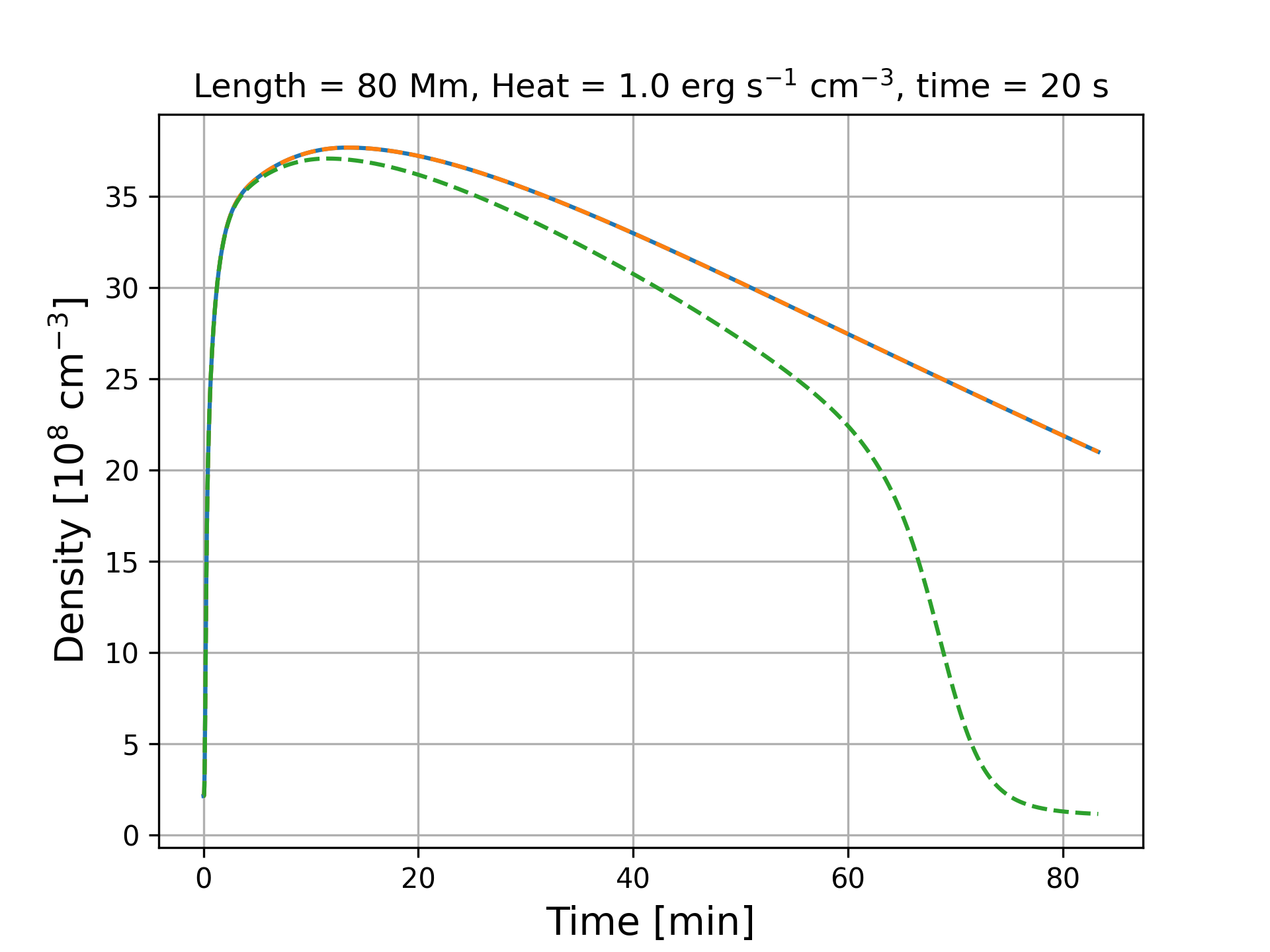}
    \includegraphics[width=0.32\linewidth]{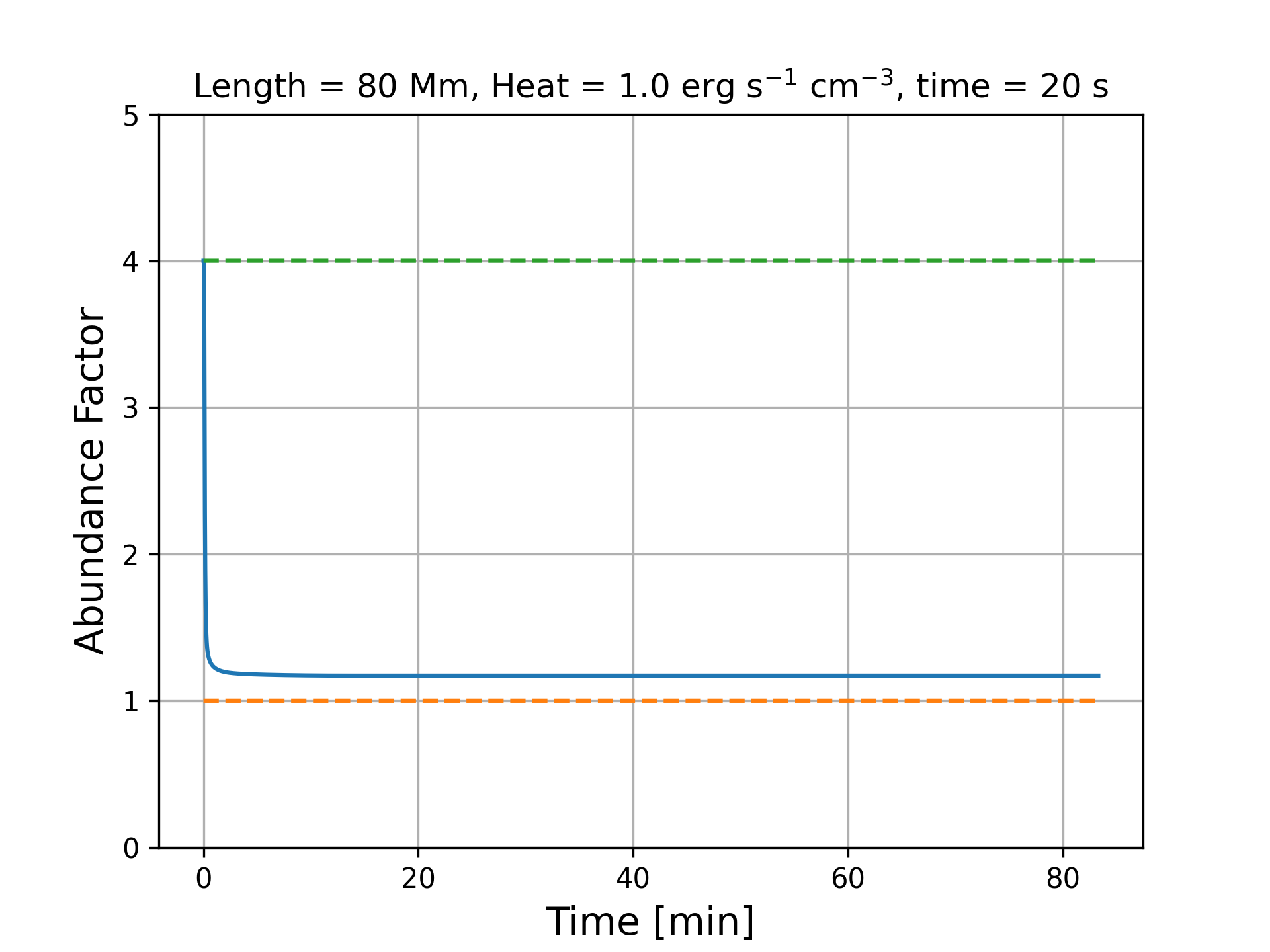}
    \caption{Similar to Figure \ref{fig:L40}, showing the results for an 80 Mm loop heated impulsively for 20 s.  \label{fig:L80}}
\end{figure*}

In Figure \ref{fig:L80}, we show a similar comparison with 80 Mm loops.  Since the cooling time depends on loop length ($\propto L^{5/6}$, \citealt{cargill1995}), we expect that the differences should be exaggerated here.  While the overall trends are similar to the previous case, there are a few differences worth noting.  The first difference is that the coronal density does not grow as large with the same heating rates, and as a consequence the abundance factor remains somewhat higher.  This is partially a consequence of the assumed form of heating, which in \texttt{ebtel++} is equivalent to a thermal conduction front.  Different heating mechanisms, such as non-thermal electron beams, would deposit energy directly in the chromosphere, which affects the flows and thus the resultant coronal densities \citep{fisher1984,allred2005,reep2015}.  The second difference is that the duration of evaporation is somewhat longer, as the flows must travel a longer distance to fill the loop, so the rate of change of the abundance factor is generally slower.  

\begin{figure*}
    \script{render_figure3.py}
    \centering
    \includegraphics[width=0.47\linewidth]{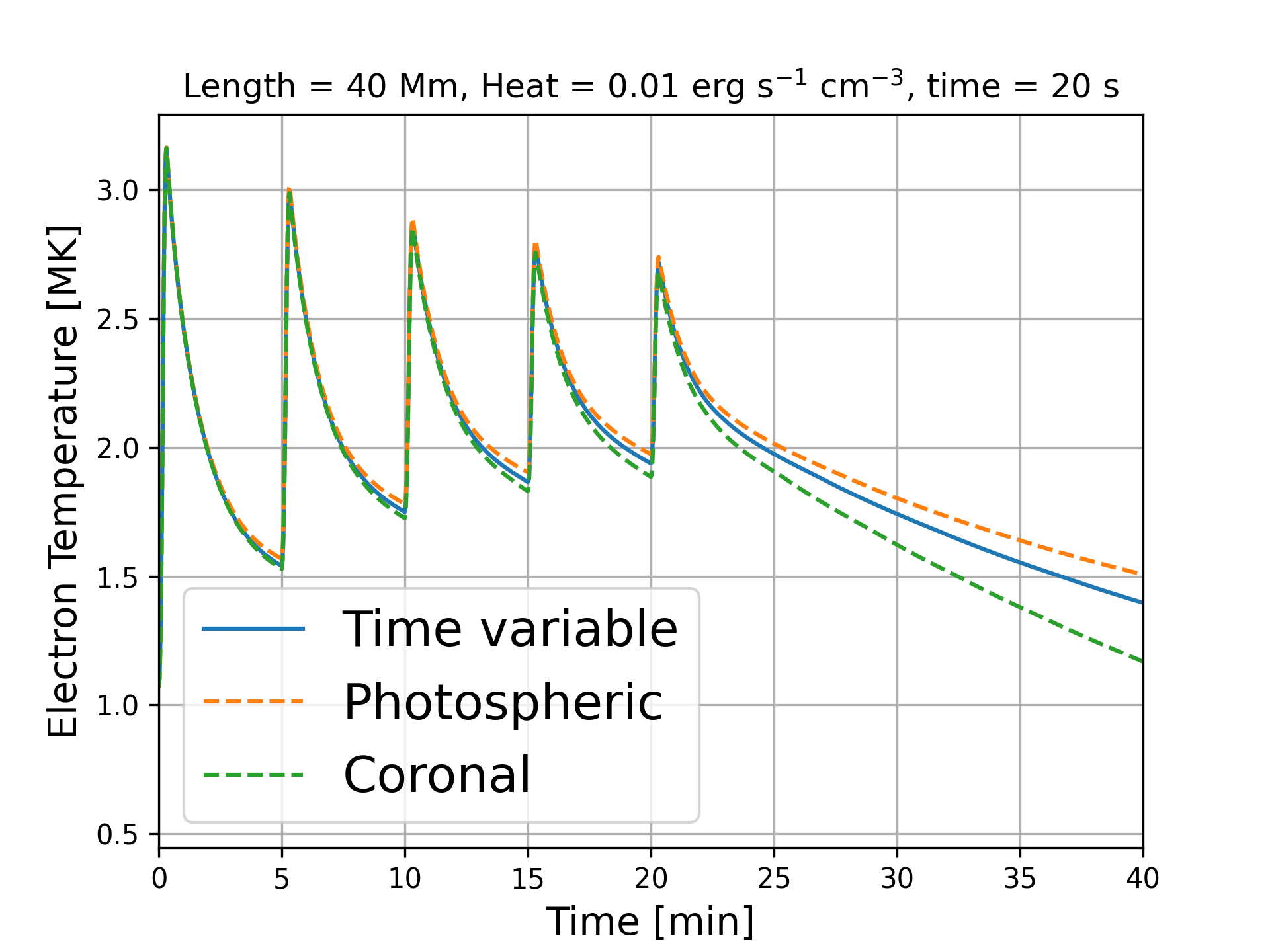}
    \includegraphics[width=0.47\linewidth]{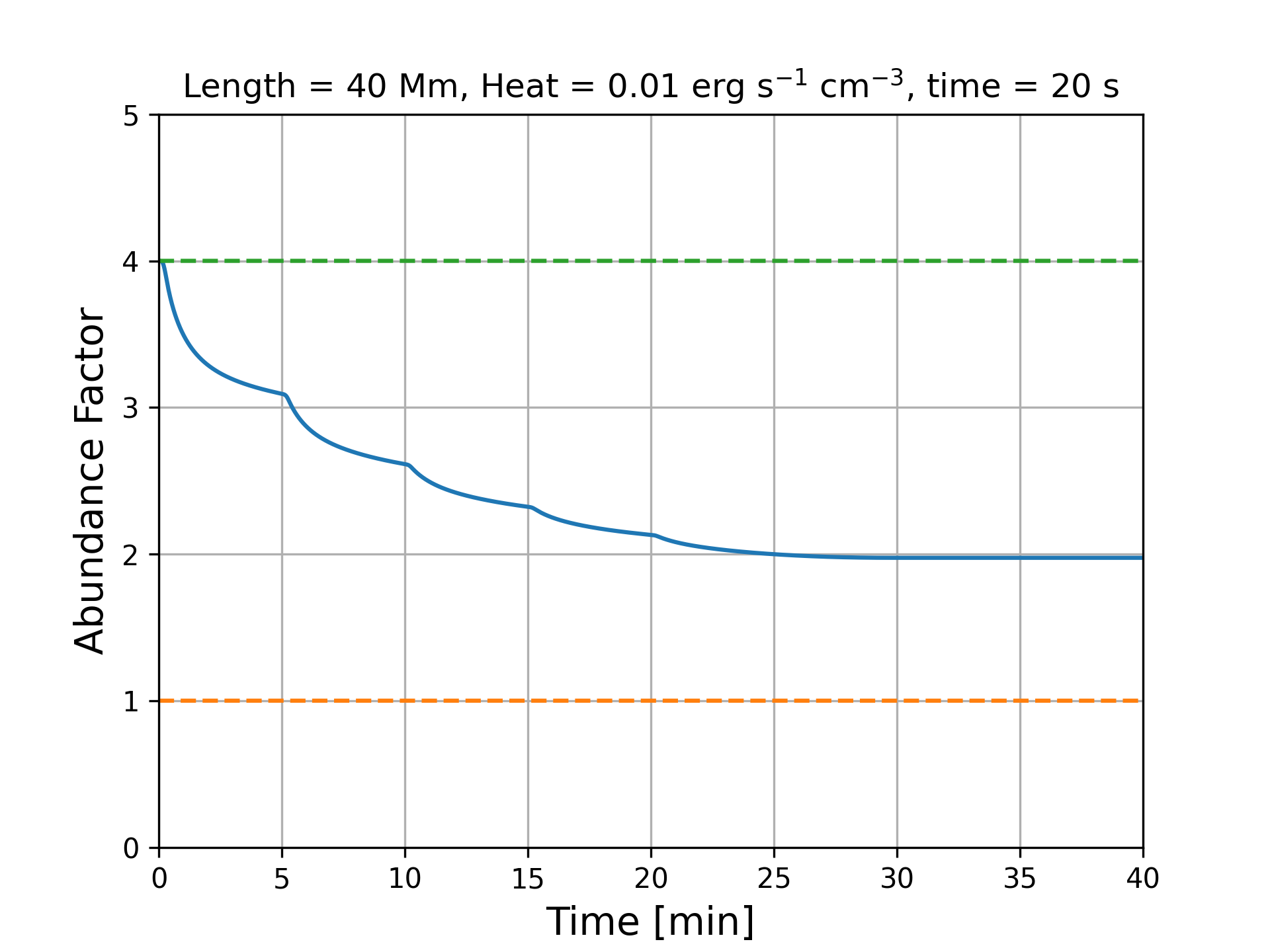}
    \includegraphics[width=0.47\linewidth]{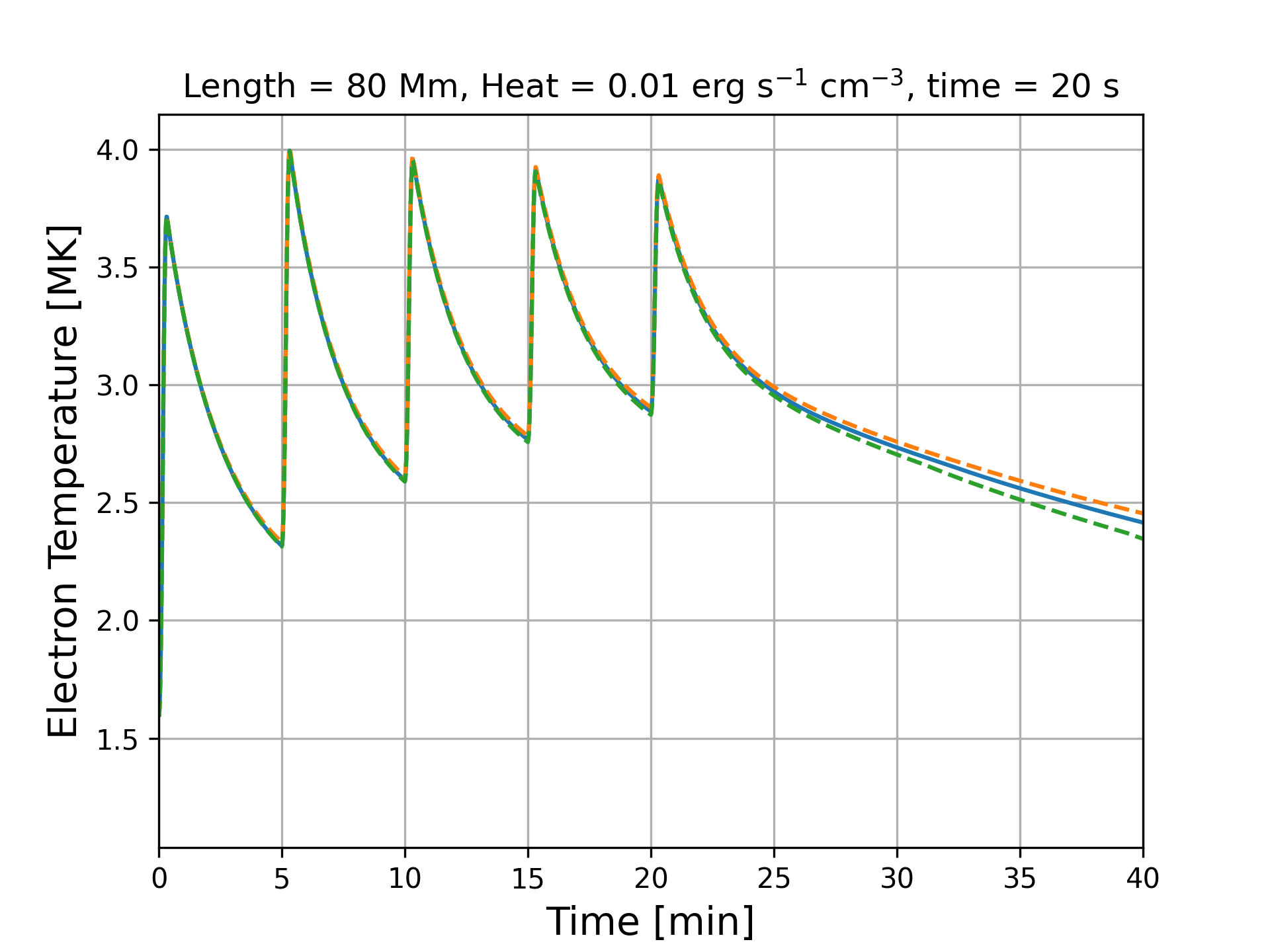}
    \includegraphics[width=0.47\linewidth]{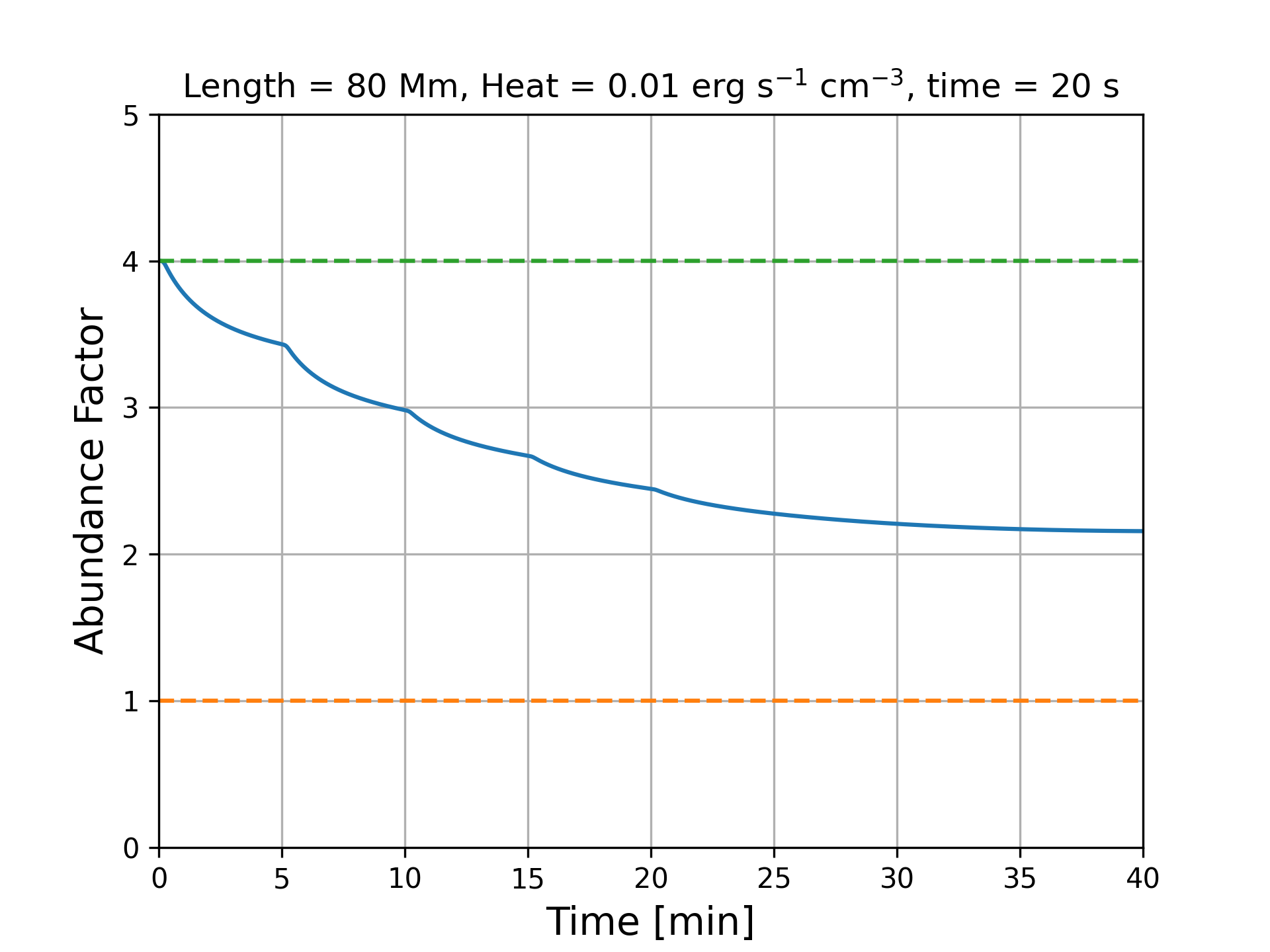}
    \caption{Two examples of nanoflare trains, with 5 heating events spaced 300 s apart, for loops of 40 and 80 Mm.  The abundance factor $f$ falls with each successive heating event. \label{fig:train}}
\end{figure*}

We finally show two examples of nanoflare trains \citep{reep2013,cargill2014,barnes2016b}, where a series of nanoflare heating events occur in close succession before ceasing and allowing the loop to cool.  Figure \ref{fig:train} shows two cases for loops of 40 and 80 Mm, with 5 heating events of 0.01 erg s$^{-1}$ cm$^{-3}$, spaced 300 s apart.  The loops oscillate around temperatures of around 2.5 and 3.5 MK, respectively, during the heating period, before rapidly cooling.  In both cases, the evaporation is prolonged for more than 30 minutes, causing the abundance factor $f$ to fall successively with each heating event (and thus evaporation event), reaching a minimum of around $f=2$.  As with single heating events, the effect of time-variable abundance on temperature and density becomes most noticeable during the cooling phase.  Of course, more realistic nanoflare trains would have stochastic waiting times, durations, and heating rates, but the abundances would similarly tend towards photospheric as each event causes evaporation.  Additionally, over long periods of time, it is likely that the Alfv\'en waves would fractionate the low FIP elements, causing a gradual return towards coronal abundances, which we have not considered here.  

\section{Discussion}

We have implemented a time-variable abundance factor $f(t)$ and radiative loss function $\Lambda(T, n, f)$ into the \texttt{ebtel++} code, in order to better understand how chromospheric evaporation changes elemental abundances, the radiative loss rate, and the cooling of coronal loops.  As evaporation carries material into the corona, the plasma there tends towards photospheric, slowing the cooling rate.  

In a typical flare loop simulation, abundances are assumed to be photospheric, as this is what has been measured in the past (\textit{e.g.} \citealt{warren2014}).  However, more recent observations from X-ray spectrometers with high cadence and high spectral resolution have shown that the abundances vary rapidly with time \citep{mondal2021}, and for (relatively) small flares, a time-fixed photospheric abundance is overly simplistic.  Recent spectroscopic observations have further found that post-flare coronal rain has photospheric abundances, while the host loop has coronal abundances \citep{brooks2024}, suggesting that the increase in density necessary to form that rain comes from material evaporated deep in the chromosphere.  

Our simulations here show that large heating rates, as might be expected in the largest flares, do drive the abundances to photospheric values almost immediately.  However, for smaller heating rates, the abundance changes (due to flows, at least) do not reach photospheric values, and therefore the cooling rate of the loop is faster.  Of course, a real flare is not composed of a single loop, so this requires a multithreaded model to further investigate the observed time-dependence.  Additionally, a higher dimensional model is required to examine the formation of coronal rain, though rain is noticeably absent in ``standard'' flare loop modeling \citep{reep2020b}.  

In an AR, the abundances can also vary with time, depending on the strength of heating and amount of fractionation by wave fluxes.  \citet{mondal2023} examined time variation of Al, Mg, Si, and S in three ARs, finding abundance factors no lower than 2 (slightly enhanced over photospheric).  Comparing the simulations here, that suggests that there may be an upper limit to nanoflare heating in those ARs of $\approx 0.1$ erg s$^{-1}$ cm$^{-3}$ since the abundances never reached photospheric values.  Furthermore, repeated heating events cause the abundance to tend towards photospheric values as the evaporation becomes prolonged, which would reduce this limit.  If the loops expand in cross-sectional area, impulsive heating causes a similarly prolonged evaporation event \citep{reep2024}, which would also push the abundances towards photospheric values and reduce this limit.

In this work, we have not attempted to include the changes in abundances due to ponderomotive forces of Alfv\'en waves propagating from the corona to chromosphere, which is thought to be the cause of the FIP effect \citep{laming2015}, where low FIP elements like iron are generally enhanced in coronal loops.  Observations of such waves have been measured (\textit{e.g.} \citealt{murabito2024}), and shown to be consistent with producing the FIP effect.  However, it is not clear what timescales are required to produce the fractionation of elements, as the modeling work to date has been time-independent.  Assuming that the time-scales are short, then this is an important effect that should additionally be incorporated to properly model time-variable abundances.  One additional complexity is that for large enough Poynting fluxes, the resistive dissipation of Alfv\'en waves in the chromosphere would also cause significant amounts of evaporation \citep{reep2016}.  Estimates of Poynting fluxes suggest that waves might carry up to $\approx 10^{10}$ erg s$^{-1}$ cm$^{-2}$ in flares \citep{russell2023}, which is large enough to drive explosive evaporation and produce photospheric abundances in general, but this is an upper limit.   The net effect on elemental abundances is not clear, therefore, and more work is required to determine the interplay.  Additionally, examination of stellar flares on stars with FIP-biased coronae (like the Sun) compared to inverse FIP-biased coronae (like HR 1099, \citealt{brinkman2001}) suggests that chromospheric evaporation is consistent with the change in abundances during the flares relative to quiescent times \citep{nordon2008}.  Further modeling work should also, therefore, examine how abundances change during stellar flares.

Finally, we have implemented this in the 0D \texttt{ebtel++} primarily as a test for its importance in loop evolution.  Higher dimensional models, however, could simulate not only the time variability but also spatial variability.  Spatial variations of abundances are observed regularly in ARs and flares \citep{doschek2018,baker2021,to2021,long2024}.  While the current model does not address this, it would be straightforward to extend this work to higher dimensions.  Since evaporation is a ubiquitous process in the solar corona, this effect is critical for (magneto)hydrodynamic simulations.

\section*{Acknowledgments}
All of the tools used to produce this paper are open-source.  The implementation into \texttt{ebtel++} is publicly available on GitHub, \url{https://github.com/rice-solar-physics/ebtelPlusPlus}, or on Zenodo, \url{https://zenodo.org/records/12675386}.

This paper is fully reproducible with the \href{https://github.com/showyourwork/showyourwork}{\showyourwork} package, and can be downloaded from \url{https://github.com/jwreep/ebtel_abundances}.  This package strives for full reproducibility and transparency in science.  Each figure in this paper contains a hyperlink with the Github logo that points to the script used to generate the plots in that figure.  Additionally, the whole paper can be generated with a single command (`showyourwork build') that will run the \texttt{ebtel++} simulations, read in the data, and produce the figures.  Users can then additionally reconfigure the simulations, the plots, or the scripts to their own liking to validate or extend this research.  The exact data used to produce the figures has been cached on Zenodo: \url{https://zenodo.org/records/12695938}.

\software{
\texttt{ebtel++}, v.0.2 \citep{barnes2024},
\texttt{matplotlib} \citep{hunter_matplotlib:_2007}, \texttt{showyourwork!} \citep{luger2021}
}

\bibliography{bib}

\end{document}